\documentclass[preprint]{elsarticle}

\newcommand{\our}{\texttt{DeepMTL}\xspace}
\newcommand{\mtl}{\texttt{MTL}\xspace}
\newcommand{\mtlss}{\texttt{MTL-SS}\xspace}
\newcommand{\mtpe}{\texttt{MTPE}\xspace}

\newcommand{\mx}{\ensuremath{\mathbf{X}}\xspace}
\newcommand{\mxx}{\ensuremath{\mathbf{X^{'}}}\xspace}
\newcommand{\mxij}{\ensuremath{\mathbf{X_{i,j}}}\xspace}
\newcommand{\mxxij}{\ensuremath{\mathbf{X^{'}_{i,j}}}\xspace}
\newcommand{\nf}{\cal{N}\xspace}
\newcommand{\imgimg}{\texttt{sen2peak}\xspace}
\newcommand{\simpeak}{\texttt{simplePeak}\xspace}
\newcommand{\yolocust}{\texttt{YOLOv3-cust}\xspace}
\newcommand{\ouryolo}{{\texttt{DeepMTL-yolo}}\xspace}
\newcommand{\ourpeak}{{\texttt{DeepMTL-peak}}\xspace}
\newcommand{\power}{{\texttt{PredPower}}\xspace}
\newcommand{\subtract}{{\texttt{SubtractNet}}\xspace}
\newcommand{\conf}{{\texttt{conf}}\xspace}
\newcommand{\nms}{{\texttt{nms}}\xspace}

\usepackage{subcaption}
\usepackage{amsmath}
\usepackage{epsfig,endnotes,color,paralist, multirow}
\usepackage{xspace}
\usepackage{esvect}
\usepackage[colorinlistoftodos,prependcaption,textsize=small]{todonotes}

\newcommand{\blue}[1]{\textcolor{blue}{#1}}

\newcommand{\bi}{\begin{itemize}}
	\newcommand{\ei}{\end{itemize} \bigskip}

\newcommand{\para}[1]{\smallskip \noindent\textbf{#1}}
\newcommand{\subpara}[1]{\smallskip \noindent \underline{#1}}
\newcommand{\eat}[1]{}

\newcommand{\softpara}[1]{\smallskip \noindent \underline{#1}}

\newcommand{\deeptx}{{\tt DeepTxFinder}\xspace}
\newcommand{\splot}{{\tt SPLOT}\xspace}
\newcommand{\splat}{{SPLAT!}\xspace}
\newcommand{\map}{\texttt{MAP$^*$}\xspace}

\newcommand{\lerr}{\mathrm{L_{err}}}
\newcommand{\mr}{\mathrm{M_{r}}}
\newcommand{\fr}{\mathrm{F_{r}}}
\newcommand{\perr}{\mathrm{P_{err}}}

\usepackage{etoolbox}
\makeatletter
\patchcmd{\@makecaption}
  {\scshape}
  {}
  {}
  {}
\makeatletter
\patchcmd{\@makecaption}
  {\\}
  {.\ }
  {}
  {}
\makeatother

\usepackage{hyperref}

\begin{document}

\begin{frontmatter}

\title{DeepMTL Pro: Deep Learning Based Multiple Transmitter Localization and Power Estimation}

\author[add1]{Caitao Zhan\corref{cor1}}
\ead{cbzhan@cs.stonybrook.edu}

\author[add1]{Mohammad Ghaderibaneh}
\ead{mghaderibane@cs.stonybrook.edu}

\author[add2]{Pranjal Sahu}
\ead{pranjal.sahu@kitware.com}

\author[add1]{Himanshu Gupta}
\ead{hgupta@cs.stonybrook.edu}

\cortext[cor1]{Corresponding author}

\affiliation[add1]{organization={Stony Brook University},
addressline={100 Nicolls Rd},
city={Stony Brook, NY},
postcode={11794},
country={USA}}

\affiliation[add2]{organization={Kitware},
addressline={101 E Eeaver St},
city={Carrboro, NC},
postcode={27510},
country={USA}}

\begin{abstract}
In this paper, we address the problem of Multiple Transmitter Localization (\mtl).
\mtl is to determine the
locations of potential multiple transmitters in a field, based on readings from a distributed set of sensors.
In contrast to the widely studied single transmitter localization problem, the \mtl problem has only been studied recently in a few works.
\mtl is of great significance in many applications wherein intruders
may be present. E.g., in shared spectrum systems, detection of unauthorized transmitters and estimating their power are imperative to 
efficient utilization of the shared spectrum.
	
In this paper, we present \our, a novel deep learning approach to address the \mtl problem.  
In particular, we frame
\mtl as a sequence of two steps, each of which is a computer vision problem: image-to-image translation and object detection. 
The first step of image-to-image translation essentially maps an input image representing sensor readings to an 
image representing the distribution of transmitter locations, and the second object detection step derives precise
locations of transmitters from the image of transmitter distributions.
For the first step, we design our learning model \imgimg, while for the second step, we customize a 
state-of-the-art object detection model \yolocust. 
Using  \our as a building block, we also develop techniques to estimate transmit power of the localized transmitters.
We demonstrate the effectiveness of our approach
via extensive large-scale simulations and show that our approach outperforms the previous approaches 
significantly (by 50\% or more) in performance metrics including localization error, miss rate, and false alarm rate. 
Our method also incurs a very small latency.
We evaluate our techniques over a small-scale area with real testbed data and the testbed results align with the simulation results.
\end{abstract}

\end{frontmatter}

\section{\bf Introduction}
\label{sec:intro}

The RF spectrum is a limited natural resource in great demand due to the
unabated increase in mobile (and hence, wireless) data
consumption~\cite{andrews2014will,sigcomm21-5G}.
In 2020, the U.S. FCC moves to free up 100 MHz of previously military occupied mid-band spectrum in the 3.45-3.55 GHz band for paving the way for 5G development. 
Also, the research and industry communities have been addressing this capacity crunch via the development of {\em shared spectrum}.
Spectrum sharing is the simultaneous usage of a specific frequency band in a specific geographical area and time by a number of independent entities where harmful electromagnetic interference is mitigated through agreement (i.e., policy, protocol)~\cite{dod20-spectrum}. 
Spectrum sharing techniques are also normally used in 5G networks to enhance spectrum efficiency~\cite{survey-specshare}.~However, protection of spectrum from unauthorized
users is important in maximizing spectrum utilization.

The increasing affordability of the software-defined radio (SDR)
technologies makes the shared spectrum particularly prone to
unauthorized usage or security attacks. With easy access to SDR
devices (e.g. HackRF, USRP), it is easy for selfish users to transmit data
on a shared spectrum without any authorization and potentially causing
harmful interference to the incumbent users.  Such illegal spectrum
usage could also happen as a result of infiltration of computer viruses
or malware on SDR devices.~\cite{survey-specshare} depicts three cases of spectrum attack.
As the fundamental objective behind such
shared spectrum paradigms is to maximize spectrum utilization, the
viability of such systems depends on the ability to effectively guard
the shared spectrum against unauthorized usage.  The current
mechanisms however to locate such unauthorized users (intruders) are
human-intensive and time-consuming, involving the FCC enforcement bureau
which detects violations via complaints and manual
investigation~\cite{mobicom17-splot}. 
Motivated by the above, we seek an effective
technique that is able to accurately localize multiple simultaneous
intruders (transmitters). Below, we describe the multiple transmitter localization problem.

\para{Multiple Transmitter Localization (\mtl).}  \eat{Localization of
unauthorized users in a shared spectrum system essentially boils down to localizing transmitters/intruders in a given area under a shared spectrum system.}
The transmitter localization problem has been well studied, but most of the focus has been on localizing a {\em single} transmitter at a time. 
However, it is important to localize multiple transmitters simultaneously to effectively guard a shared spectrum
system. E.g., a malware or virus-based attachment could simultaneously 
cause many devices to violate spectrum allocation rules; spectrum
jamming attacks would typically involve multiple transmitters. More
importantly, a technique limited by the localization of a single intruder
could then be easily circumvented by an offender by using multiple
devices. The key challenge in solving the multiple transmitter localization (\mtl) 
problem comes from the fact that
the deployed sensor would receive only a {\em sum} of the signals from multiple transmitters, and separating the signals may be impossible. 
\eat{In
addition, the other challenge that \mtl in the context of shared
spectrum system poses is the presence of authorized users---e.g., the
incumbent users and the dynamic set of secondary users that have been
allocated spectrum by the manager.}

\softpara{Prior Works.}
The \mtl problem has been recently addressed in a few prior works, among which \splot~\cite{mobicom17-splot}, \map~\cite{ipsn20-mtl}, and \deeptx ~\cite{icccn20-deeptxfinder} are the most 
prominent. \splot essentially decomposes the \mtl
problem to multiple single-transmitter localization problems based on
the sensors with the highest power readings in a
neighborhood. However, their technique implicitly assumes a propagation model, and
thus, may not work effectively in areas with complex propagation
characteristics, and it is not effective in the case of transmitters
being located close by (a key challenging scenario for \mtl problem).
Our recent work \map solves the \mtl problem using a 
hypothesis-driven Bayesian approach; in particular, it uses prior training in the form of distributions
of sensor readings for various transmitter locations, and uses the training data to determine
the most likely configuration (i.e., transmitters' locations and powers) for a 
given vector of sensor readings. 
However, to circumvent the high computational cost of a pure Bayesian approach,
\map uses a divide and conquer heuristic which results in somewhat 
high number of misses and false alarms while still incurring high
latency. 
\deeptx uses a CNN-based learning model approach; 
however, they use a separate CNN model for a specific number of transmitters
and thus may incur high model complexity and training cost while also limiting the number 
of transmitters that can be localized.
In our evaluations, we compare our work with each of the above approaches.

\begin{figure}[t]
\centering
\includegraphics[width=0.75\textwidth]{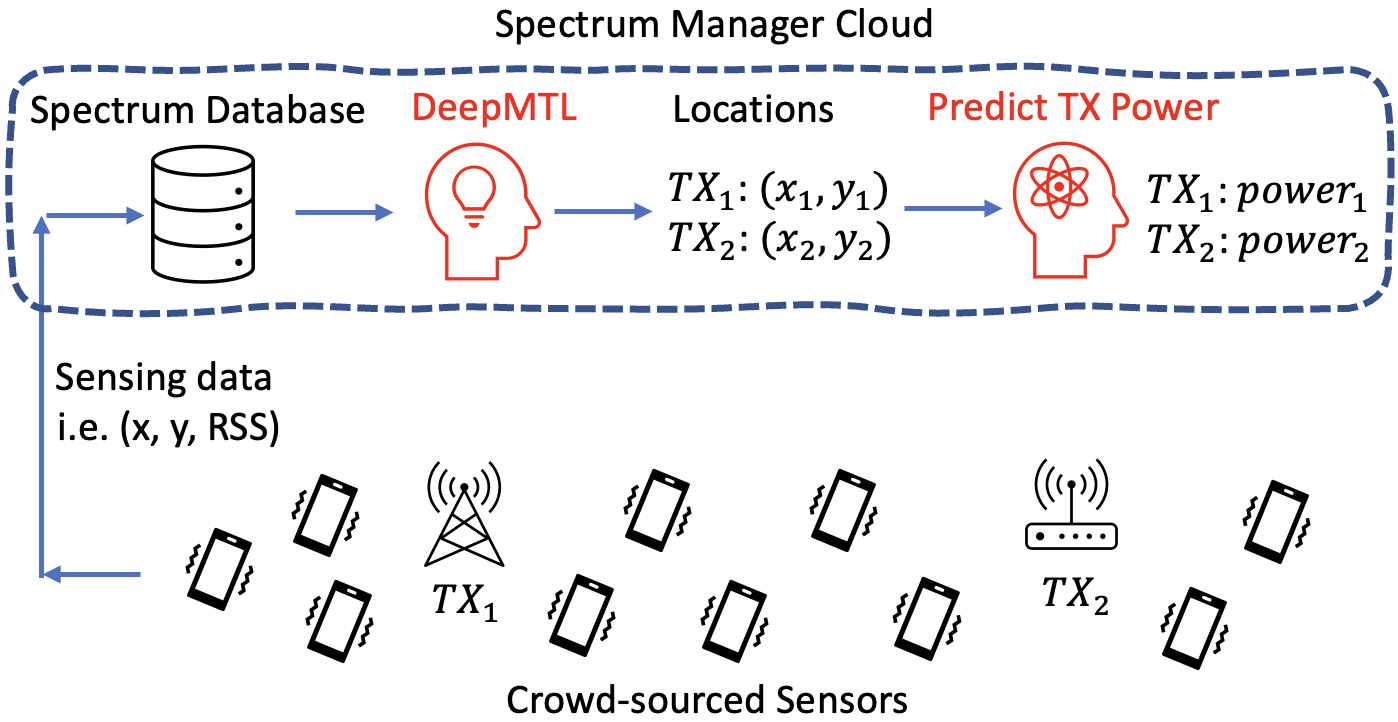}
\vspace{-0.1in}
\caption{Multiple transmitter localization using a distributed set of sensors. Sensing data is uploaded to a spectrum manager server in the cloud. \our is a deep learning approach to multiple transmitter localization which helps protect 
spectrum against unauthorized usage. After that, prediction of transmission powers happens using \our as a building block.}
\vspace{-0.15in}
\label{fig:illustration}
\end{figure}

\para{\our: Our Two-Step Approach.}  As in prior
works~\cite{mobicom17-splot,chakraborty2017specsense}, we assume a
crowdsourced sensing architecture (See Fig.~\ref{fig:illustration}) wherein relatively low-cost spectrum
sensors are available for gathering signal strength in the form of received power.
We use a convolutional neural network (CNN) based approach
to solve the \mtl problem. In particular, we frame
\mtl as a sequence of two steps: image-to-image translation and object detection, each of which
is solved using a trained CNN model. 
The first step of image-to-image translation maps an input image representing sensor readings to an 
image representing the distribution of transmitter locations, and the second object detection step derives precise
locations of transmitters from the image of transmitter distributions.
We name our \mtl approach as \our.

\softpara{Motivation.}
Our overall approach and its various aspects are motivated by the following considerations. 
{\bf First}, we use a learning-based strategy to 
preclude assuming a propagation model~\cite{mobicom17-splot} or conducting surveys of sensors reading distributions~\cite{ipsn20-mtl}.
Assumption of propagation model suffers from the fact that 
even sophisticated propagation models yield unsatisfactory accuracy and thus lead to degraded performance.
Among all learning-based strategies, deep learning can implicitly capture the environment characteristics (e.g., objects, walls, landscape) in the neural network layers' weights learned through the training of the data~\cite{mobicom20-deeploc}.
Even though a learning-based approach incurs a one-time high training cost,
it generally incurs minimal latency during inference, which is an important consideration for our \mtl problem.
The intruder detection should incur minimal latency to be effective.
{\bf Second}, the geographical nature of the \mtl problem suggests that convolutional neural networks (CNNs) are 
well-suited for efficient
learning of the desired function. In particular, the features of the \mtl problem can be
represented in an image (2D matrix) corresponding to their geographic locations, which can be fed as an input
to an appropriate CNN model which can leverage the spatial correlation among the input features 
to facilitate efficient learning. 
{\bf Lastly,} we use a two-step architecture to  facilitate efficient training by essentially
providing an additional intermediate image. 
In particular,  we are able to map each step to well-studied standard computer vision problems, allowing us to build upon known techniques. 

\para{Overall Contributions.}  The goal of our work is to develop an
efficient technique for accurate localization of simultaneously
present multiple transmitters/intruders. We also extend our technique to address various
extensions such as power estimation and the presence of authorized users. Overall, we make the following 
contributions.
\begin{enumerate}
\item
For the \mtl problem, we develop a novel two-step CNN-based approach called \our approach. 
For the first step of image-to-image translation, we develop a CNN model that translates an image representing the sensor readings into an intermediate image that encodes distributions of transmitter locations (Section \ref{sec:translate}). 
For the second step of mapping transmitter distributions to precision locations via object detection,
we customize the well-known  object detection method YOLOv3 (Section \ref{sec:detect}).

\item
For localization of transmitters in presence of authorized users, we 
augment the \our model by adding a pre-processing step based on a 
CNN-model that first reduces the 
sensor readings by the power received from the authorized users (Section \ref{sec:authorized}).

\item
To estimate transmit power of the intruders, we augment our \our model
with a power-estimation CNN-model which iteratively estimates the power of transmitters
in sub-areas (Section \ref{sec:power}).

\item
We evaluate our techniques via large-scale simulations as well as a small-scale testbed data and demonstrate their effectiveness and superior
performance compared to the prior works (Section \ref{sec:evaluation}).
\end{enumerate}

A preliminary version of this paper appeared at IEEE WoWMoM 2021~\cite{wowmom21-deepmtl}.

\section{\bf Background, \mtl Problem and Our Approach}
\label{sec:problem}
\label{sec:prob-def}

In this section, we describe the background of the shared spectrum systems,
formulate the \mtl problem, then describe our methodology. 

\para{Shared Spectrum System.} In a shared spectrum paradigm, the
spectrum is shared among licensed users (primary users, PUs) and
unlicensed users (secondary users, SUs) in such a way that the
transmission from secondaries does not interfere with that of the
primaries (or secondaries from a higher-tier, in case of a multi-tier
shared spectrum system). In some shared spectrum systems,
the location and transmit power of the primary users may be
unavailable, as is the case with military or navy radars in the CBRS band.
Such sharing of spectrum is generally orchestrated by a centralized
entity called {\em spectrum manager}, such as a spectrum
database in TV white
space~\cite{sas-paper} or a central spectrum access system in
the CBRS 3.5GHz shared band~\cite{milind2015dyspan}. The spectrum
manager allocates spectrum to requesting secondaries (i.e., permission
to transmit up to a certain transmit power at their location) appropriately
so as to avoid interference to primaries.
Users that transmit without explicit permission are referred to as 
unauthorized users or {\em intruders}; the \mtl problem is to essentially
localize such intruders. 

\para{\mtl Problem.}  Consider a geographic
area with a shared spectrum. Without loss of generality, we assume a
single wireless frequency\footnote{To avoid confusion with image channels, we use {\em wireless frequency} instead of the perhaps more appropriate {\em wireless channel} term.} throughout this paper\footnote{Multiple wireless frequencies can be handled independently. 
Note that if we assume the wireless propagation characteristics to be similar for different frequencies, then we do not need to train different models for each of them. Our localization techniques would still work
for scenarios wherein the intruders may change their transmit frequencies dynamically.}.
For localization of intruders, we assume
available crowdsourced sensors that can observe received signal in the wireless frequency of interest, and compute (total) received signal strength (RSS).
RSS can be measured using low-cost sensors and has been shown to achieve good accuracy for single-transmitter localization~\cite{infocom00-radar}.
In the related work Section~\ref{sec:related}, we will discuss signal metrics other than RSS, such as AoA, ToA, etc.~At any instant, there may be a set of intruders
present in the area with each intruder at a certain location transmitting
with a certain power which may be different for different intruders.

The \mtl problem is to determine the set of intruders with their
locations at each instant of time, based on the set of sensor
observations at that instant. 
For the main 
\mtl problem, we assume that there are no primary or authorized users, and thus, assume that the
sensor readings represent aggregate received power from the transmitters we
wish to localize.
However, in Section \ref{sec:authorized}, we investigate the more general \mtl problem where the background primary and/or secondary users may also be present.

\begin{figure*}
    \centering
    \includegraphics[width=\textwidth]{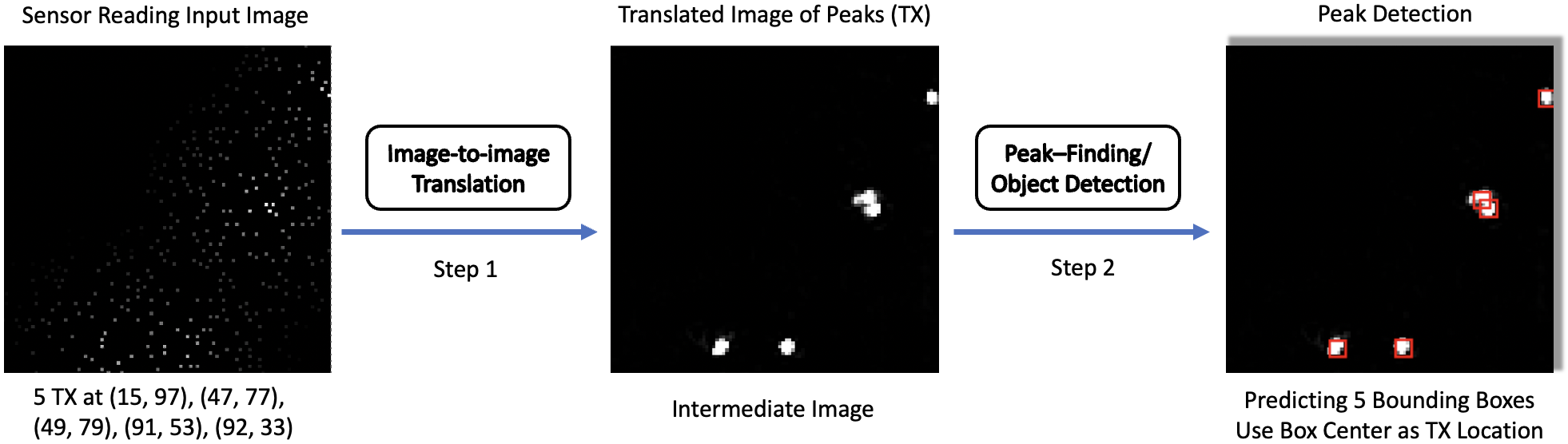}
    \vspace{-0.1in}
    \caption{The overall two-step CNN architecture of the \our model. The first step is the \imgimg, whose higher idea is to translate the input image of sensor readings to the image of peaks where each peak implies a transmitter. The \imgimg architecture is illustrated in Fig.~\ref{fig:part1}. The second step is \yolocust, a customized version of YOLOv3, to perform object/peak detection in the output image of the first step. This step returns the precise location coordinates of TX. The \yolocust architecture is illustrated in Fig.~\ref{fig:yolo}. A zoom-in of the peak detection result of the second step is in Fig.~\ref{fig:peaks}.}
    \label{fig:overall}
\end{figure*}

\para{\bf Our Approach.}
In our context, each sensor communicates its observation to a centralized spectrum 
manager which then runs localization algorithms to localize any potential (multiple) transmitters. 
We design and implement a novel two-step localization algorithm named \our, as illustrated in Fig.~\ref{fig:overall}, based on CNN models. 
The first step (Section \ref{sec:translate}) is a four-layer image-to-image translation CNN model that is trained
to translate an input image representing sensor readings to an image of 
transmitters' locations distributions. Each distribution of a transmitter can be visualized as a mountain with a peak, so we name this model \imgimg.
The second step (Section \ref{sec:detect}), called \yolocust, is a customized object-detection method build upon YOLOv3\cite{yolov3} which localize the objects/peaks in the translated image.
The high-level motivation behind our two-step design is to frame the overall \mtl
problem in terms of well-studied learning problem(s). 
The two steps facilitate efficient learning of the models by supplying an intermediate image with the training samples. 

\section{\bf \our Step 1: Sensor Readings to TX Location Distributions}
\label{sec:translate}

In this section, we present the first step of our overall approach to the \mtl problem, i.e., 
the image-to-image translation step which translates/transforms the sensor reading to distributions
of TX locations. Here, we first create a grayscale image to represent the input sensor readings;
this image encodes both the sensors' RSS readings and the sensors' physical location. We then train and use
a convolutional neural network (CNN) model to transform this input image
to an output image which represents the distribution of TX locations. Pixels in the output image that have higher values will have a higher chance of having a TX being present at that location. 
 
\para{Input/Output Image Sizes and Tiling Approach for Large Areas.}
We need to represent data by images of certain sizes.
Typically, an image should be a size of a few hundred pixels by a few hundred pixels, since a thousand pixels by thousand pixels images will consume too much GPU memory.
In this paper, we pick $100\times 100$ as the size for both our input and output images in the first image-to-image translation step.
Given an area that we want to monitor and a $100\times100$ size image, we will know how large an area a pixel will represent and we call it a pixel subarea.
A large pixel subarea could certainly lead to high localization errors, due
to very coarse granularity. We can address this by using a ``tiling"
technique, wherein we divide the given area into tiles, then represent each 
tile by $100\times100$ size image and use our localization techniques in the tile.
We can do some post-processing to handle cross-tiling issues (e.g., \cite{icccn20-deeptxfinder} uses overlapping tiles and employs a voting scheme inside the overlapping tile area).
 
\eat{\blue{For our \our, each pixel represents a $10m\times 10m$ area. Since \our input image is of size $100\times 100$, the whole input image represents a $1000m\times1000m$ area. 
To deal with a larger area, let's say $10000m\times 10000m$ area, two simple methods are: increase the input image size to $1000\times 1000$; 
Or let each pixel represent a $100m\times 100m$ area.
The first method requires redesigning \our (changing layer hyperparameter etc.).
Note that the input image size is usually in the range of lower hundreds. 
For example, $224\times 224$ is used by ResNet \cite{resnet} and $416\times 416$ is used by YOLO. 
$1000\times 1000$ is generally speaking too big as it will consume too much GPU memory.
Also, a pixel representing a larger area means coarser granularity and lower accuracy.
A reasonable approach will be keeping the \our as it is, and use a ``sliding window" like method, i.e. slide the \our 100 times (10 rows, 10 columns) to cover the $10000m\times10000m$ area.
\deeptx\cite{icccn20-deeptxfinder} is using this idea and they name it a tile-based approach.
To deal with smaller areas, we can fill the rest of the area with noise floor values. 
By filling, we can make smaller images fit into the $100\times100$ input size requirement of \imgimg.
In the following subsection \ref{subsec:ipsn}, we demonstrate how we generate a $100\times 100$ size image from a $10\times 10$ size grid in real-world testbed data.}
}


\subsection{\bf Input Image Representing Sensors' Readings}

We localize transmitters based on observations from a set of sensors, i.e. solve the \mtl problem assuming only intruders.
The input of the localization method is sensor observations. 
Here, an {\em observation} at a sensor is the received power (RSS, in decibels) over a time window of a certain duration, in the frequency of interest (we assume only one wireless frequency). 
RSS is computed using FFT over the I/Q samples collected in a time window. More specifically, in our evaluations, we use a Python API \cite{psd} that computes the power spectral density from a sequence of signal data (I/Q samples), and then, we choose the RSS at the frequency of interest.
\begin{figure}[t]
    \centering
    \includegraphics[width=0.75\textwidth]{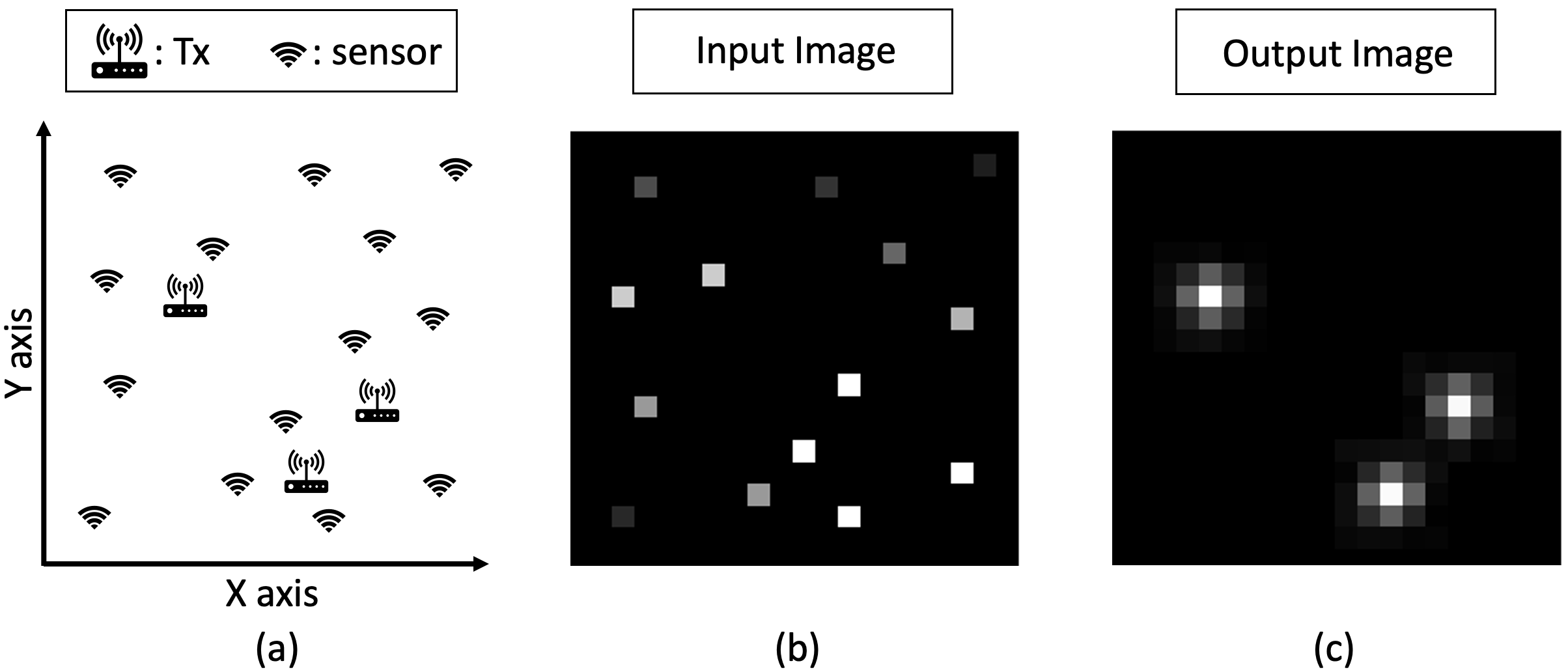}
    \vspace{-0.1in}
    \caption{Illustration of \our first step's input and output images. (a) Area with distributed sensors and transmitters to be localized. (b) Input image representing the sensor readings (RSS) and locations. (c) Output Image, where we put a 2D Gaussian distribution with its ``peak" at the transmitter's location.}
    \label{fig:input}
\end{figure}
Different than~\cite{mobicom17-splot,ipsn20-mtl}, we represent the sensor information, i.e.,
their locations and observations, in a 2D input image.
We use a 2D grayscale image, and let us denote it $\mx$. The pixel \mxij denotes the observation of the sensor at the grid cell whose index is $(i, j)$. For example, $\mx_{10, 20}=-50$ denotes there is a sensor at coordinate $(10, 20)$ with an RSS reading of $-50$ dB. 
If there is no sensor at location $(i, j)$, we assign the noise floor $\nf$ (i.e. -80 dB) value to \mxij.
Note that the above pixel values (representing the sensor observations) are not the standard
image pixel values that lie in the [0, 255] range.
Also, since the pathloss computed by propagation models during simulations could be real numbers, the sensor observation values could be real numbers. 
So we use a 2D matrix with real numbers instead of an image object.

Before passing this sensor reading image as input to our CNN model, we do a normalization step; we first subtract the $\nf$ from each value and then divide it by $-\nf$/2.
Let $\mxx$ denote the 2D matrix after the normalization of $\mx$. 
The value $\mxxij$ will be zero at locations without sensors, and \mxxij will be a positive real number (in most cases, less than two) for locations with sensors. E.g., if $\mx_{10, 20}=-50$, then the 
$\mxx_{10, 20}$ equals to $(-50- (-80)) / 40) = 0.75$.
Fig.~\ref{fig:input}~(b) shows how a matrix is used to represent the input information that contains both the RSS and the spatial location of the distributed sensors in an area that exists 14 sensors in Fig.~\ref{fig:input}(a).



\subsection{\bf Output Image Representing TX locations' Distributions}
\label{subsec:imgimg_output_image}
We now focus on designing the output image to represent the distribution of TX locations; 
the output image is essentially the ``label" assigned to each input image that guides the training of the CNN model. 
Fig.~\ref{fig:input}(c) illustrates the output image of the image-to-image translation step in Fig.~\ref{fig:input}(a) that contains three transmitters.

A straightforward representation that represents the TXs with locations is to just
use an array of $(x, y)$ elements where each $(x, y)$ element is the location of a 
transmitter, as in~\cite{icccn20-deeptxfinder}.
However, this simple representation is less conducive to efficient model learning,
as the representation moves away from spatial representation (by representing locations 
as positions in the image) to direct representation of locations by coordinate values.
E.g., in~\cite{icccn20-deeptxfinder}'s CNN-based approach to \mtl problem, the authors
assume a maximum number $N$ of transmitters and train as many as $N+2$ different CNN models
and thus, limiting the overall solution to the pre-defined maximum number of 
transmitters.
Instead, in our approach, we facilitate the learning of the overall model, by solving the
\mtl problem in two steps, and in this step of translating sensors' reading
to transmitter locations' distributions, we represent the output also as an image.
This approach allows us to use a spatial learning model (e.g. CNN) for the second step too, and preclude
use of regression or fully-connected layers in the first step. 

Inspired by recent work on wireless localization problem~\cite{mobicom20-deeploc} that represents the input and output as images, we represent our output of the first step as an image as well.
The output image is a grayscale image implemented as a 2D matrix with real numbers. 
In the output image, we use 25 ($5\times 5$) pixel values to represent the presence of a transmitter. 
It is desirable to use an odd side length square (e.g., $3\times 3$, $5\times5$, $7\times 7$) for symmetry. 
For a $100\times 100$ size input we use, while $3\times3$ gives too little information for a transmitter and $7\times 7$ generates too many overlaps for close by transmitters, $5\times 5$ is the sweet spot.
Other pixels far away from any transmitter are zero-valued.
Among multiple potential ways to represent a transmitter presence by a number of pixels, we found
that using a 2D Gaussian distribution around the pixel of TX location, as shown in Fig.~\ref{fig:input}(c), yields the best model performance.
Thus, a geographic area with multiple transmitters present is represented by a grayscale image with multiple Gaussian distributions, with each Gaussian distribution's peak inside the pixel corresponding to transmitter's location. 
Based on preliminary performance
tests, we pick the amplitude of the 2D Gaussian peak to 10, the standard deviation to 0.9, and located the center of the distribution at the location of each transmitter.
Note that the location of the TX is in continuous domain and usually not at the center of the grid cell.



\begin{figure}
    \centering
    \includegraphics[width=\textwidth]{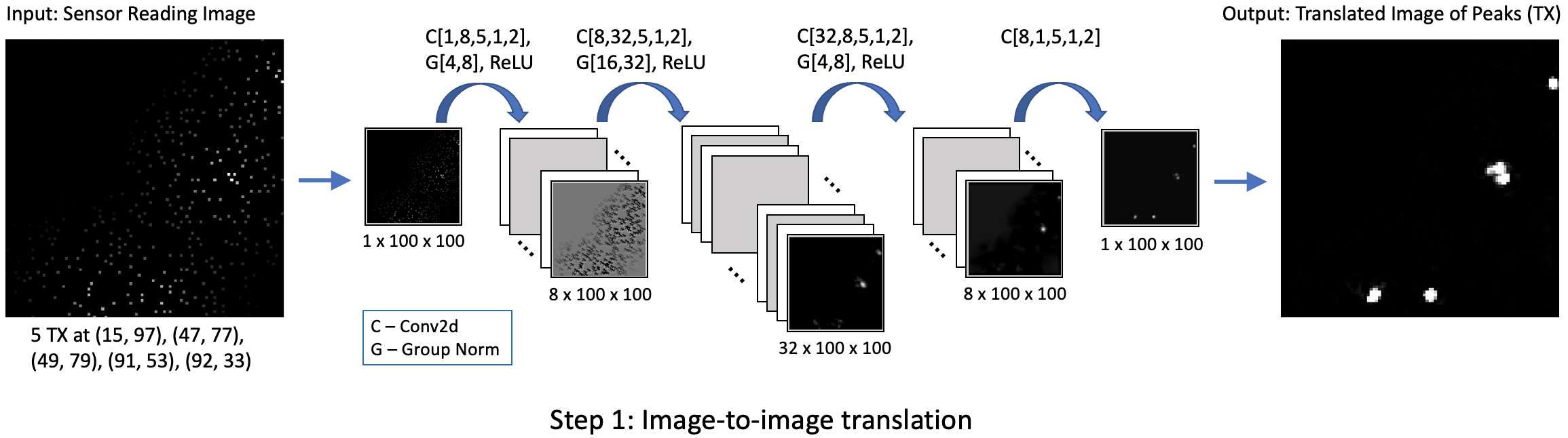}
    \vspace{-0.1in}
    \caption{Architecture of the first step CNN, a four layer image-to-image translation model (\imgimg). The figure displays how the data volume flows through the various convolutional layers. C stands for Conv2d, and for each Conv2d layer, the five values shown are [number of input channels, number of output channels, kernel size, stride, padding]. G stands for group normalization, and, for each group normalization, the two values shown are [number of groups, number of channels]. See~\S\ref{sec:translate} for details.}
    \label{fig:part1}
\end{figure}

\subsection{\bf Image-to-Image Translation: \imgimg CNN Model}
\label{subsec:img-translate}

At a higher level, we use a deep and spatial neural network, in particular a CNN, to learn the
approximation function that maps the input image (of sensor readings) to the output image (of Gaussian distributions for TX locations). We refer to this as the {\em image-to-image translation}
model. Our approach is inspired by the recent work~\cite{mobicom20-deeploc} that frames 
a different wireless localization problem as an image-to-image translation problem.
We incorporate the idea into our multiple transmitter localization problem and utilize recent advances in the computer vision area. 
Encoder-decoder based CNN models like U-Net~\cite{miccai15-unet} with down-sampling
and up-sampling convolutional layers have been successful in effectively learning image-to-image translation functions. However, in our setting, we observe that the usage of down-sampling layers (such as max-pooling) degrades the performance of the model, especially in the case when transmitters may be close to each other wherein the model is unable to distinguish the nearby transmitters and generate a single large distribution in the output image. To circumvent this, we avoid using any
down-sampling layers in our model and redesign the image-to-image translation model as described below.


\para{\imgimg CNN Model}. 
We refer to our image-to-image translation CNN model as \imgimg, as it translates sensors' 
readings to ``peaks" with Gaussian distributions corresponding to transmitter locations.  
It has four \footnote{We observe that a four-layer lightweight and symmetric \imgimg  model produces good results and adding more layers gives marginal improvement.}
convolutional layers, as shown in Fig.~\ref{fig:overall}(a). 
We use an input size of $100\times100$. The number of convolutional filters are varying for
different layers, with up to 32 in one of the layers. 
We tried doubling the filter numbers at each layer, but it does not lead to significant 
improvement (it does yield a lower error, but the output image does not improve significantly
to impact the second step of our architecture). We use a kernel size of $5\times5$, a stride of 1, and a padding of 2.
This ensures that the dimensions do not decrease and all the pixels are treated 
uniformly, including the ones at the edge of the image.
With the above four convolutional layers, the receptive field~\cite{receptive-field} of each neuron in the output layer is $17\times17$.
Normalization layers can improve the learning process. We chose group normalization \cite{groupnorm} and put it after the first three convolutional layers. 
We compared group and batch normalization~\cite{batchnorm} methods in our context, and observed
better performance with the group normalization. 
For the activation layers, we select rectified linear unit (ReLU) and put it after the group normalization layers.


\softpara{The Loss Function}. Our inputs ($X$) and output ($Y$) are images. We use L2 loss function which computes the mean squared error aggregated over individual pixels.
More formally, our loss function is  defined as:
\begin{equation}
 \frac{1}{N} \sum_{i}^{N} || \imgimg(X_i) - Y_i ||^2
 \label{equ:sen2peak_loss}
\end{equation}
where $N$ is the number of samples used in computing the loss, $|| \cdot ||^2$ is L2 loss function, $X_i$ and $Y_i$ are the $i_{th}$ sample's input and output images respectively, and $\imgimg(X_i)$ is the predicted output image corresponding to the input $X_i$. During training, we use 
Adam~\cite{kingma2017adam} as the optimizer that minimizes the loss function. We set the learning rate to 0.001 and the number of epochs to 20 and the model converges well.

\section{\bf \our Step 2: TX Locations' Distributions to Precise Locations}
\label{sec:detect}

In this section, we present the second step of our overall localization approach. We refer to
this step as the {\em peak detection} step, as the goal is to detect the peaks within the 
Gaussian distributions in the input image (which is also the output image of the first step).
The first step outputs an image that has multiple distributions (presumably, Gaussian), whose
peaks need to be interpreted as precise locations of the transmitters/intruders.
As, our end goal is to determine the precise locations of the present transmitters, we develop techniques to detect peaks within  the output image of the first step. 
We propose two different strategies for the peak-detection task. The first strategy is a straightforward peak detection algorithm based on finding local maximal values, while the 
second strategy is based on framing the problem as an object detection task; for the second strategy, we utilize a widely used state-of-the-art computer vision model called YOLOv3~\cite{yolov3}.

\para{Simple Peak Detection Method.}
The simple and straightforward peak detection method is to designate pixels with
locally maximal values as peaks, subject to certain thresholds.
More formally, we use a threshold $x$ for a peak value, and also use a parameter $r$ to define a $r$-radius neighborhood of a pixel. 
Then, any pixel whose value is more than $x$ and is the maximum among all pixels with a $r$-radius neighborhood, is designated as a peak (transmitter location). 
We use $x=2$ and $r=3$, in our evaluations.
Note that each pixel represents a subarea; thus, a pixel designated as pixel only 
implies the transmitter location at the {\em center} of the corresponding subarea.
To localize the transmission more precisely with the pixel's subarea, we use a scheme that localizes the transmitter within the subarea by computing a weighted average of the peak pixel's coordinate and the peak's neighbor pixels' coordinates.
The weight of a pixel is the predicted pixel value itself from the first step \imgimg.
We refer to the above simple approach for the second-step of \our as \simpeak.

\begin{figure}
	\centering
	\includegraphics[width=0.75\textwidth]{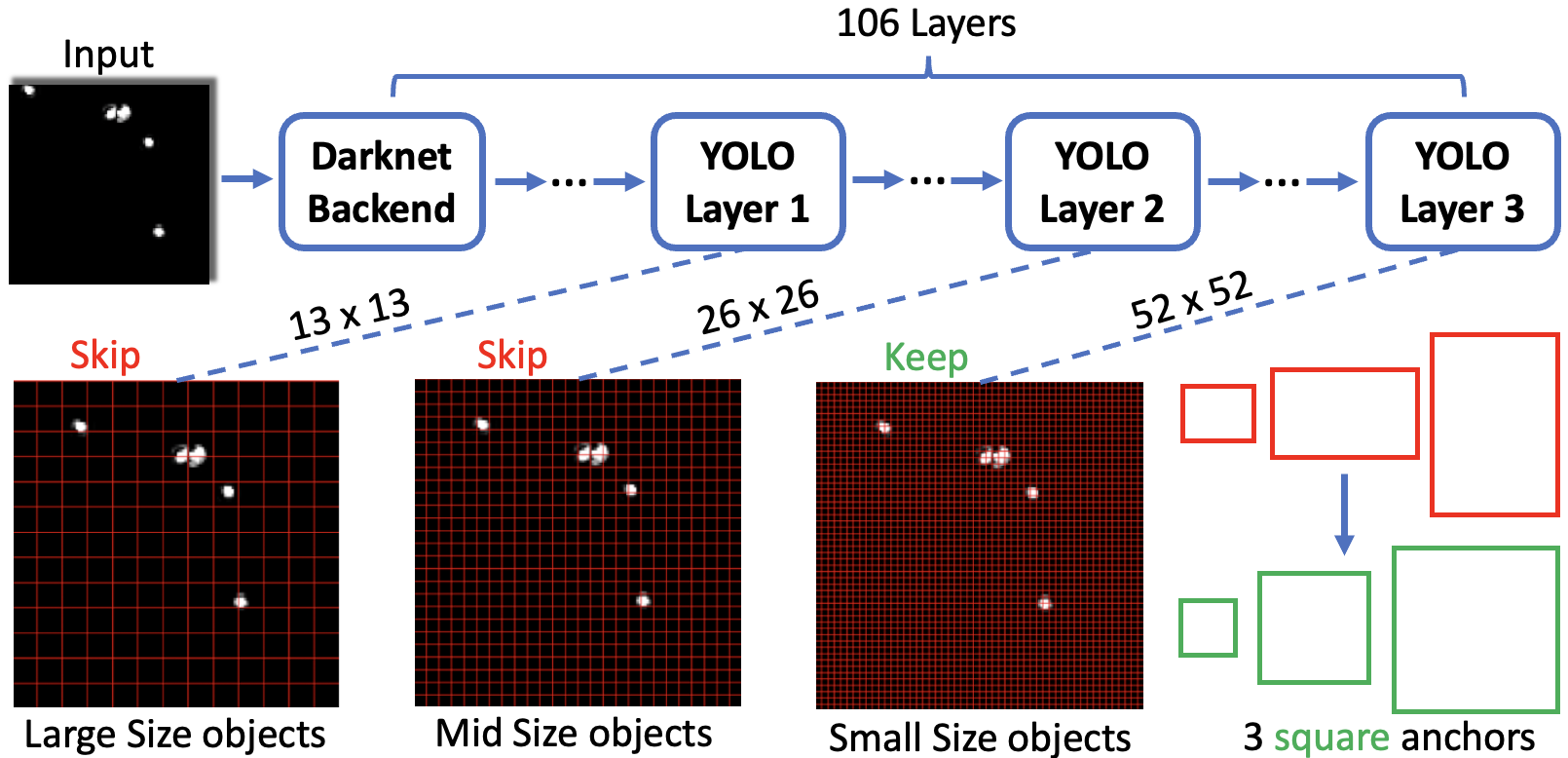}
	\vspace{-0.1in}
	\caption{Our \yolocust in the second step of the \our. The two major customization are: (i) Use only the third YOLO layer that detects small size objects (the output of \yolocust is the bounding box predicted by the third YOLO layer and we use the center of the bounding box as the transmitter location), and (ii) change the rectangle anchors to square anchors.}
	\label{fig:yolo}
\end{figure}

\subsection{\bf Object-Detection Based Precise Localization: \yolocust} 

The simple hand-crafted method described in the previous subsection performs reasonable well in most cases in our simulations. 
However, its key drawback is that it needs appropriate threshold values that may vary from case to case; such thresholds can be difficult to determine, especially since
the input images (with distributions) are not expected to be perfect as they are themselves output of a learning model.
Inaccurate threshold values can lead to false alarms and misses. 
Also, the previous method is not sufficiently accurate at the sub-pixel level, where
each pixel may represent a large area such as $10m \times 10m$ or even $100m \times 100m$. Thus,
we propose a CNN-based learning method that overcomes the above shortcomings.  
CNN has been widely used for object detection in different areas~\cite{objectdetectionsurvey,alizadeh21}.

\begin{figure}[t]
	\centering
	\includegraphics[width=0.75\textwidth]{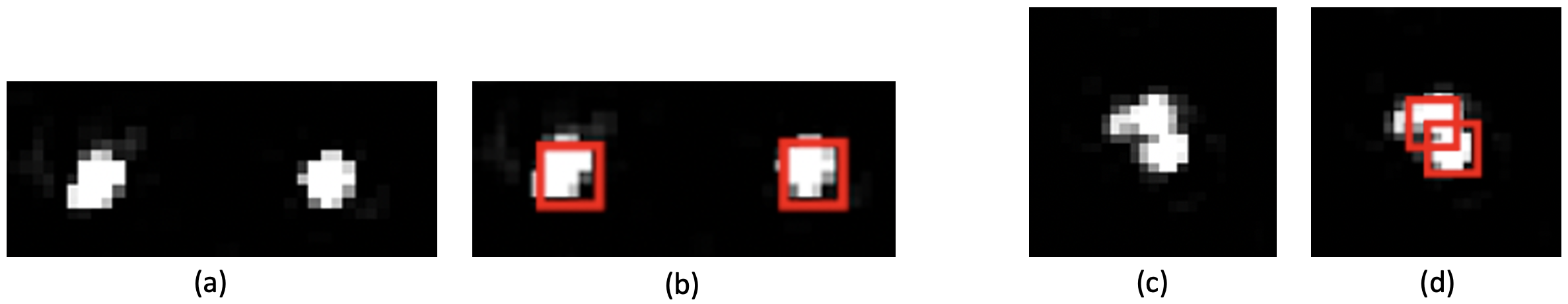}
	\vspace{-0.1in}
	\caption{(a) is the zoom-in of two peaks at the bottom of the Fig.~\ref{fig:overall} example.
	(c) is the zoom-in of the two close by peaks in the middle right of the Fig.~\ref{fig:overall} example. (b) and (d) shows the bounding boxes that \yolocust outputs for (a) and (c) respectively.}
	\label{fig:peaks}
\end{figure}

We frame this problem as an object detection task where the objective is to detect and localize
known objects in a given image. We observe that our second-step peak detection problem is essentially an object detection problem where the ``object" to detect is a ``peak".
Thus, we turn the \mtl problem of localizing multiple transmitters into detecting 
peaks in the images output by \imgimg model. 
For object/peak detection, we design \yolocust, our customized version of YOLOv3~\cite{yolov3}.
Fig.~\ref{fig:peaks} is a zoom-in of localizing two close by transmitters (peaks) in Fig.~\ref{fig:overall}(b).

\softpara{Peak Detection Using \yolocust.} 
Object detectors are usually comprised of two parts: (i) a backbone which is usually pre-trained on ImageNet, and (ii) a front part (head), which is used to predict bounding boxes of objects, probability of an object present, and the  object class. 
For the front part, object detectors are usually classified into two categories, i.e., one-stage detectors such as the YOLO~\cite{cvpr16-yolo} series, and two-stage detectors such as the R-CNN~\cite{cvpr14-rcnn} series.
We choose the one-stage YOLO series because of its computational efficiency, high popularity and available ways to customize it for our specific context. We refer to the customized version as \yolocust, see Fig.~\ref{fig:yolo}.
Implementing a 106-layer deep neural network with a complex design from scratch is
out of scope of our work. 
Thus, we use a publicly available source repository~\cite{yolo-github} and made customization on top of it.
We refer to the architecture that uses \imgimg $\ $and \yolocust in sequence as \our, our key product.  
In addition, we use \imgimg in combination with the uncustomized original YOLOv3,
and refer to it as \ouryolo (still change the class number to one).

\softpara{Customization of YOLOv3}.
Overall, we incorporated four customization to YOLOv3, of which two are significant and the
other two are relatively minor. See Table~\ref{table:yolocust}. YOLOv3 is designed to be a general object detector that can detect objects of various sizes, shapes, and classes within input images
of various sizes. However, in our context, the input images are of a fixed size, with
only a single class of objects which are relatively small and semi-circular. 
Based on the above observations, we make changes to the original YOLOv3 that both 
decrease the model complexity and improve its performance.

\begin{table}[t]
    \caption{Differences between the original YOLOv3 and our \yolocust.}
    \centering
    \begin{tabular}{ |p{5cm}|p{5cm}| } 
     \hline
     YOLOv3 & \yolocust  \\
     \hline \hline
     Has three YOLO layers at 13x13, 26x26, and 52x52 for detection & Only use the last 52x52 YOLO layer for detection (skip the first two YOLO layers) \\
     \hline
     Has 3 different rectangle anchors for each YOLO layer & Has 3 square anchors \\
     \hline
     Every 10 batches, randomly chooses a new input image dimension size & Do not randomly choose new input dimension size \\ 
     \hline
     Has 80 different categories of object class & Only has one category for the peak class \\ 
     \hline
    \end{tabular}
    \label{table:yolocust}
\end{table}

{\em Customization Details}. 
The first and second changes presented in Table~\ref{table:yolocust} are major changes and we elaborate them in the following paragraphs.
Making prediction at three different scales is one of the highlights of YOLOv3 and an improvement comparing to the previous version YOLOv2 which was prone to missing at detecting small objects. 
As shown in Fig.~\ref{fig:yolo}, the coarse-grain $13\times13$ YOLO layer-1 is designed for detecting large size objects, 
the $26\times 26$ YOLO layer-2 is designed for detecting middle-sized objects, and 
the fine-grained $52\times52$ YOLO layer-3 is designed for detecting small-sized objects.
Since the peaks in our translated images are always small objects, we only use the last $52\times 52$ YOLO 
detection layer (and skip the first two YOLO layers).
As shown in Fig.~\ref{fig:yolo}, by ``skipping" the two YOLO layers means that we do not use them in 
computing the overall loss function and their outputs are not used in predicting the bounding boxes.
In our \yolocust, the only YOLO layer predicts 8112 bounding boxes, since it has a 
dimension of $52\times 52$ and each cell results in prediction of 3 bounding boxes; this is in contrast
to the original YOLOv3, which predicts 10647 bounding boxes ($3 \times (13\times13 + 26\times26 + 52\times52) = 10647$).

The anchor box is one of the most important hyperparameters of YOLOv3 that can be tuned to improve
its performance on a given dataset.  The original YOLO's anchor boxes are $10\times13$,  $16\times30$, and $33\times23$ (for
the input image of size $416\times416$ pixels), which are essentially bounding boxes of a rectangular shape. 
These original YOLOv3 anchors were designed for the Microsoft COCO \cite{mscoco} data set, and were chosen
since they best describe the dimensions of the real world objects in the MS COCO data set. In our context,
since the peaks are generally squares---we use the anchor boxes to be $15\times15$, $25\times25$, and $35\times35$.

\begin{figure}[t]
	\centering
	\includegraphics[width=0.9\textwidth]{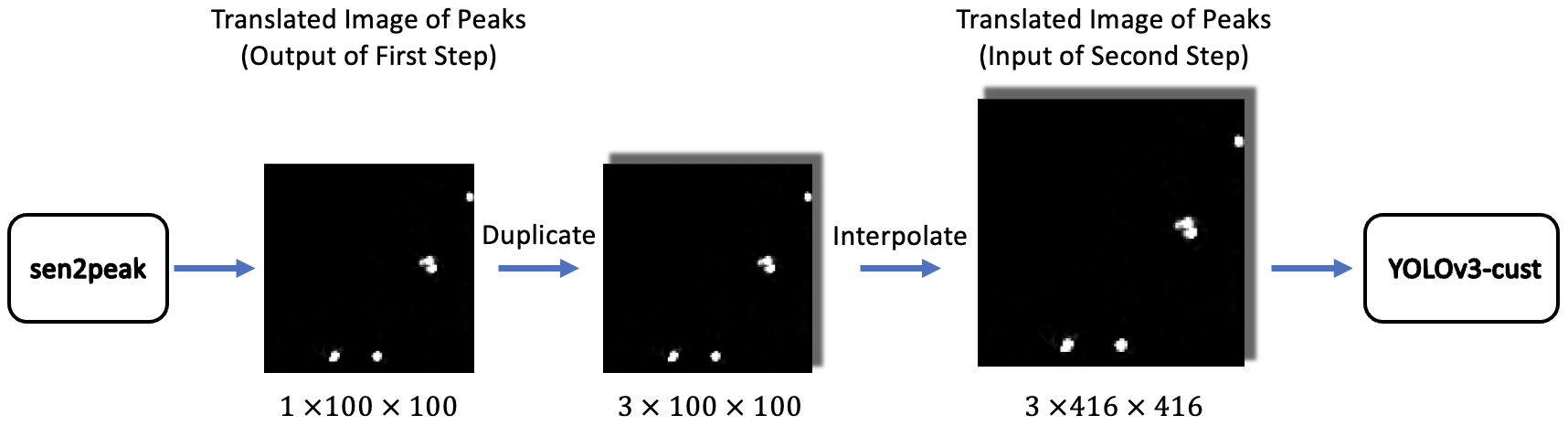}
	\vspace{-0.1in}
	\caption{The data processing of \imgimg's output to get \yolocust's input of correct size.}
	\label{fig:yolo-preprocess}
\end{figure}

\softpara{Input Image for \yolocust.}
The first step \imgimg's output image is $100\times100$, while the second step \yolocust's input is required\footnote{YOLOv3 was developed before our work and the YOLOv3 authors set the input size of the CNN model to $3\times416\times416$.~Although we are customizing their YOLOv3 model, we cannot change the input size because changing it will change the convolutional layer structure, which will 
preclude us from using the pre-trained weights in the YOLOv3 backbone.} 
to be a three-channel (RGB) image with each channel being size of $416\times 416$. 
To feed the output of \imgimg to \yolocust, we do the following: (i) First, we duplicate 
the \imgimg's output image to create two more copies and thus create a three-channel image
of 100 $\times$ 100 size channels; (ii) Next, we resize the $100\times100$ channels to $416\times416$ channels using the PyTorch's default ``nearest neighbor" interpolation. See Fig.~\ref{fig:yolo-preprocess}.

\softpara{Output of \yolocust.}
YOLO treats objected detection as a regression problem. The regression target (or ``label") for an object is a five-value tuple $(x, y,$ $length,$ $width, class)$. 
In our case, there is only one $class$. 
$x$ and $y$ are real number location coordinates of the center of the bounding box, which we use as the location of the transmitter. 
$Width$ and $height$ determine the size and shape of the object---which we consistently set to be 5 each to signify a $5\times5$ square. 
Note that the center of the bounding box is in the continuous domain. 
Thus, we are able to get sub-pixel level location of the transmitters.

\section{Localization in the Presence of Authorized Users}
\label{sec:authorized}

Till now, we have assumed that the only transmitters present in the area are the intruders which need to be localized. In this section, we solve the more general \mtl problem, where there may be a set of authorized users in the background. 
This is referred to as the multiple transmitter localization - shared spectrum (\mtlss) problem \cite{ipsn20-mtl}.

\begin{figure}
    \centering
    \includegraphics[width=\textwidth]{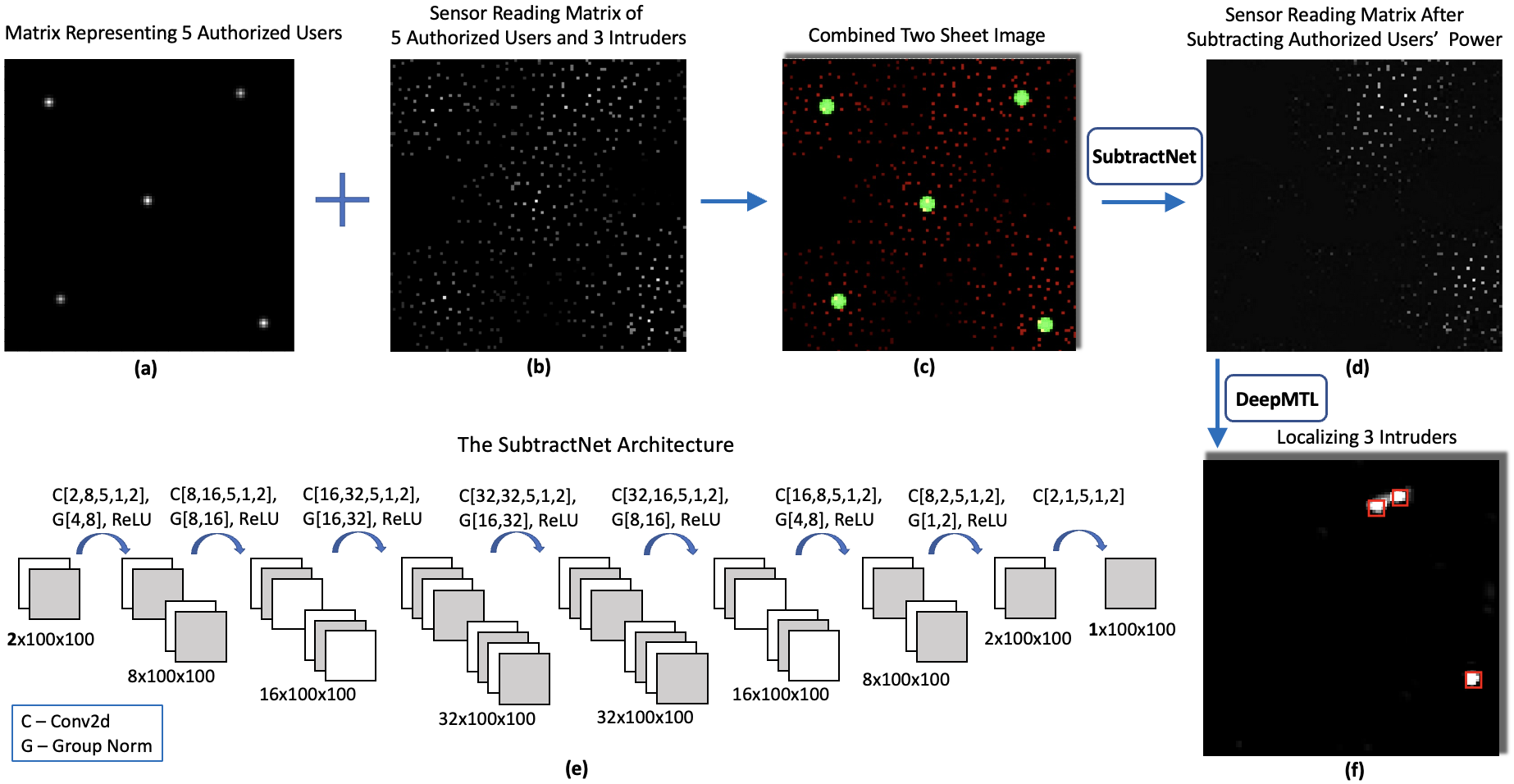}
    \vspace{-0.1in}
    \caption{Overall architecture of second approach to localize 3 intruders in the presence of 5 authorized users. The input of the \subtract is (c), which is stacking authorized user matrix (a) and the sensor reading matrix (b). (d) is the output of \subtract, where the transmission power of the authorized users is subtracted from the area. The details of the \subtract model is in (e). (f) is the localization output after feeding (d) into \our.}
    \label{fig:subtractnet}
\end{figure}

In particular, in a shared spectrum paradigm, there are primary users and an evolving set of active secondary users transmitting in the background.
Different than the intruders whose locations are unknown, the authorized users' locations are known and we wish to utilize this known information to better localize the unknown intruders.
The key challenges come from the fact that the set of authorized users is not static and changes over time as allocation requests are granted and/or active secondary users become inactive over time.
A straightforward way to handle background authorized users is to localize every transmitter, and then remove the authorized users. 
However, any localization approach is susceptible to performance degradation with the increase in the number of transmitters to be localized.
Thus, the straightforward approach of localizing every transmitter is likely to be error-prone.
Therefore, we attempt to develop a new approach that uses \our as a building block that uses the information of the location of the authorized uses in a way other than removing them after localizing all.
The new approach tries to subtract the received signal strength at the sensors by a value received from the authorized users.
This subtraction is done by a novel CNN model; we refer to it as \subtract.
Then we feed the image with subtracted powers to the \our and get the locations of the intruders.
See Fig.~\ref{fig:subtractnet}(c)--(d)--(f).
We describe \subtract in the following paragraphs.


\softpara{\subtract Input Image}. The sensor reading has two sources, one is the intruders and the other is the authorized users.
We aim to subtract the power of the authorized users and remain the power from the intruders.
So the input of the \subtract will contain two kinds of information: the authorized users' information (Fig.~\ref{fig:subtractnet}(a)), including both the location and the transmitter power, and the sensor reading matrix (Fig.~\ref{fig:subtractnet}(b)) that encode the power from all transmitters.
To incorporate the two kinds of information, we first encode the authorized user information into a matrix that has the same dimension as the sensor reading matrix.
Then stack the two matrices together.
The combined stacked image is nothing but a two-channel image, which can be interpreted as Red and Green channels.
The sensor reading matrix is the Red channel and the authorized user matrix is the Green channel. There is no Blue channel.
To represent the authorized transmitter in the Green channel, we use a Gaussian peak similar to what we did in the \imgimg for representing transmitters (Section \ref{sec:translate}).
The difference is that in \imgimg, all the peaks have a uniform height, whereas in \subtract, the height of the peak is the power of the authorized transmitter.
So the higher the power of the authorized transmitter, the higher the peak in the Green channel.
Another difference is that the authorized transmitters are approximated at discrete locations instead of the continuous locations as in \imgimg.

\softpara{\subtract Output Image}. 
The \subtract's output image is just a one-channel images and represents the sensor readings due to the intruders only.  

\softpara{\subtract CNN Architecture}.
We refer to the model that subtracts the power from the authorized users as the \subtract. 
It has a similar design philosophy with \imgimg.
\subtract is also an image-to-image translation neural network.
Compared to \imgimg, it doubled the number of layers, mainly because \subtract needs a bigger receptive field than \imgimg.
A bigger receptive field can let the CNN model update sensors that are further away from the authorized user.
For the loss function, we use the L2 loss function, similar to the loss function used in Equation~\ref{equ:sen2peak_loss}, merely replacing the \imgimg with \subtract in Equation~\ref{equ:sen2peak_loss}.
The training details are also the same as in \imgimg.

\section{Estimating the Transmit Power of Transmitters}
\label{sec:power}

In this section, we extend our techniques to estimate the transmit power of the
intruders; we refer to the overall problem as  Multiple Transmitter Power Estimation (\mtpe). Estimation of the transmit power of transmitters can be very useful in the shared
spectrum systems. In particular, estimated transmit powers of the primary users (if unknown, as in the case of military users or legacy systems) can be used to set a
``protective" region around them---inside which secondary users can be disallowed~\cite{Ureten2011powerlocation}.
Estimating transmit power of secondary users can also be useful. E.g., if the violation
in a shared spectrum system is based on a certain minimum threshold, then it is important to estimate the transmit power to determine a violation. 
Also, the estimated transmit power of secondary users can also be used to ``circumvent" their intrusion---i.e., for the primary users to appropriately increase their transmit power to overcome the harmful interference from the secondary users. 
In general, estimating the transmission power is beneficial to various operations such as node localization, event classification, jammer detection \cite{PowerEstimate2010Zafer}.
 
There are several works that estimates the transmission power of a single transmitter, often jointly with its location \cite{PowerEstimate2010Zafer, Ureten2011powerlocation, icoin2007-powerposition}.
Our previous work \cite{ipsn20-mtl} can estimate the power of multiple transmitters.
The similarity among all four of these methods is that they are estimating the power and location jointly.
In this paper, we propose a new method that leverages the capabilities of \our by using it as a building block. We first localize the transmitters by \our. 
Then given the localized locations, estimate the transmitters' transmission power by a newly designed CNN model \power. 
Although \power is designed to only estimate the power of a single transmitter, we use it together with a machine learning-based error correction method that can mitigate the errors while applying \power to the multiple transmitter power estimation scenario.

In this section, we develop a technique to predict the transmission powers of the intruders. Here, for simplicity, we assume no background authorized users, though, the techniques in this section also work in the presence of authorized users. 
We leverage our accurate and robust localization solver that tolerates varying transmission power for different transmitters (the varying transmission power needs to be in a range).
We propose an efficient approach and its overall methodology at a high-level is as follows.
And then in the next subsection we describe our \power model.
\begin{enumerate}
\item We use \our to localize the multiple transmitters in a field. 
\item We develop a CNN model \power to predict power of a single isolated (far away from other intruders) intruder. 
\item For other (non-isolated) intruders, we still use \power to predict their powers but employ a post-processing ``correction" technique to account for nearby intruders. 
\end{enumerate}

\subsection{\bf \power: Predicting Power of a Single Isolated TX}
\label{subsec:in-out-design}

\softpara{\power Input Image}. 
Let us consider an ``isolated" transmitter $T$. 
To predict $T$'s power, we start with
creating a smaller-size image by cropping the original sensor readings image with the area of a certain size around $T$. 
In our evaluations in Section \ref{sec:evaluation}, the transmitters have a transmit radius\footnote{I.e., sensors beyond a distance of 20 pixels away from a transmitter $x$ receive only negligible power from $x$.}
of around 20 pixels, which is equivalent to 200 meters.\footnote{Transmission ranges of a standard 2.4 GHz and 5 GHz WiFi at default transmission powers (100 mW) are roughly 45m and 15m respectively. 
In our simulations (Section \ref{sec:evaluation}), we use the 600 MHz frequency band. 
As the lower the signal frequency, the higher the transmission range, a transmission range of around 200m is reasonable.}
For this setting, we used an cropped area of $21 \times 21$ around the isolated transmitter $T$ to predict its power, with $T$ is at the center of this area; also, in this setting, we define a transmitter to be {\em isolated} if there is no other transmitter within a 20-pixel 
distance.\footnote{Ideally, transmitters with a transmit radius of 20 pixels should entail defining isolated transmitters as ones that have no other transmitters within a 40-pixel distance, and then use a
$41\times41$ area around the isolated transmitter. However, in our evaluations, our chosen values yielded a more efficient technique with sufficient accuracy.}
Note that the above cropping process requires the location of the transmitter to be known, and hence, we undertake the above power-estimation process after the localization of the transmitters using the \our model.  
We crop images from the same dataset where \our is trained on.

\softpara{\power Output Power}.
The output of the \power is a single pixel whose value is the predicted power of the transmitter located at the center of the cropped image.
Before coming into this single pixel output design, we tried using the height or radius of the peak from the output of \imgimg to indicate the power. 
But we figure out that the height or radius of the peak is hard to accurately predict and therefore is not an accurate indicator of the power.
\eat{We also tried to borrow the density map idea from the crowd counting literature~\cite{ECCV18-crowdcount,aaai21-topocrowdcount} from the computer vision community.
In crowd counting solutions, the CNN model predicts a density map and the summation of the value of all the pixels is the number of crowd, i.e., the number of persons in an image.
We tried to create a concept called power density map so that the summation of the pixel values equals to the power of a transmitter.
But it didn't work out well either.
In the end, we figured out that directly predicting a floating point scalar value is the right way.
The other ways are just doing thing indirectly and making the CNN model more complex and harder to train.}
So we reduced the output complexity and designed the output as a simple single pixel whose value directly represents the power of the transmitter.
By simplifying both the input side and output side, we can design and implement a novel CNN model that can accurately predict the power of a single transmitter, as described in the following paragraph.

\begin{figure}
    \centering
    \includegraphics[width=\textwidth]{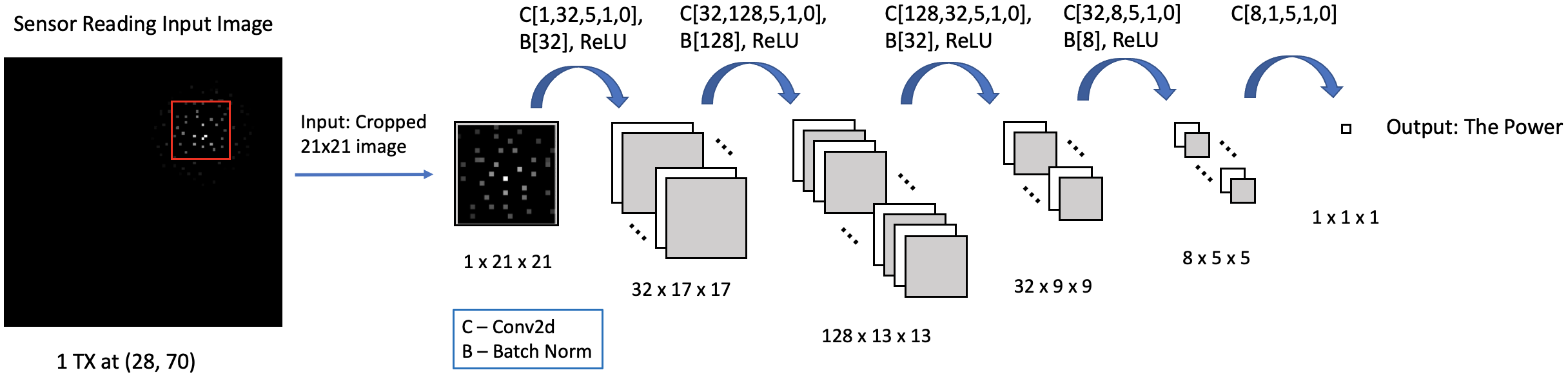}
    \vspace{-0.15in}
    \caption{Architecture of the \power, a five-layer CNN model that takes in a cropped image from the original input image and outputs the predicted power of one transmitter. The figure displays how the data volume flows through the various convolutional layers. C stands for Conv2d, a 2D convolutional layer, and for each Conv2d layer, the five values shown are [number of input channel, number of output channel, kernel size, stride, padding]. B stands for batch normalization 2d, and for each batch normalization, the value shown is [number-of-features].}
    \label{fig:power_predict}
\end{figure}

\softpara{\power CNN Architecture.}
We refer to our CNN model that estimates the power of a single transmitter as \power.
See Fig.~\ref{fig:power_predict}.
It has a similar design to \imgimg as well, where it has no max-pooling layers and no fully connected layers.
We do not use the fully connected layers and design a fully-convolutional network since the usage of fully connected layers will destroy the spatial relationships.
\power has five CNN layers and each CNN layer has a kernel size $5\times5$, striding 1 and padding 0.
With this setting, a pixel in the output layer has a receptive field of $21\times 21$, which is exactly the size of the input cropped image. Also note that the pixel is exactly at the location where the transmitter is assumed to be located (recall that the transmitter is at the center of the cropped image).
We tried both batch normalization and group normalization and found that batch normalization is better than group normalization, which is the opposite to the \imgimg scenario.
ReLU is used as the activation function.

{\em Loss Function.} The output of the last convolutional layer is technically a 3D cube, although $1 \times 1 \times 1$. So we flatten it in the end to get one scalar value.
We use a L2 loss function, which is formally defined as:
\begin{equation}
 \frac{1}{N} \sum_{i}^{N} (\power(X^c_{i}) - y_i)^2,
\end{equation}
where $N$ is the number of training samples,~$X^c_{i}$ is the cropped input image for the $i^{th}$ sample and $y_i$ is the ground truth power for the $i^{th}$ sample.
$\power(X^c_{i})$ is the predicted power.
We use Adam as the optimizer, and set the learning rate to 0.001 and the number of epochs to 20, which is sufficient for the model convergence.

\subsection{ \bf Estimating Powers of Multiple Transmitters}
Our end goal is to estimate the power of multiple transmitters at the same time. 
When the multiple transmitters are far away and isolated from each other, the problem reduces to single transmitter power estimation, which \power handles well.
The hard part is to estimate transmit powers of multiple transmitters that are close by. 
In this case, a sensor will receive an aggregated power from multiple transmitters. We assume that blind source power separation is not viable. 

\subpara{Overall High-Level Approach}.
For each localized intruder by using \our (whether isolated or not), we crop the $21\times 21$ size area around it and feed it to \power, and estimate its power. If it is actually isolated, then the predicted power is final. If it is not isolated, then we apply a post-processing correction phase to account for the overestimation of the powers, as described below.

\softpara{Correction Method for Close by Transmitters}.
Let us first consider the case where there are two close by transmitters $T_0$ and $T_1$. We use \power to estimate the power of two transmitters and get $p_0^{'}$ and $p_1^{'}$ respectively.
Let us say the ground truth are $p_0$ and $p_1$ respectively.
The estimated power will most likely be higher than the ground true power, i.e., $p_0^{'} > p_0$ and $p_1^{'} > p_1$.
Because \power can only ``see" one transmitter, and it will view two transmitters in the areas as a combined single one.
Let us focus on $T_0$  and assume $\delta_0 = p_0^{'} - p_0$.
The intuition is that $\delta_0$ has some underlying patterns that we are able to recognize.
We model $\delta_0$ as a function of some features related to $T_0$ and $T_1$.
We model $\delta_0$ as follows,
\begin{equation}
  \delta_0 =   \theta_0 \cdot p_0^{'} + \theta_{(1,1)} \cdot d_{01} + \theta_{(1,2)} \cdot p_1^{'} + \theta_{(1,3)} \cdot \frac{p_1^{'}}{d_{01}} 
  \label{equ:twotxpower}
\end{equation}
where $d_{01}$ is the distance between $T_0$ and $T_1$, and the four $\theta$s are the coefficients for the four terms respectively.
The first term is related to $T_0$ itself, and the other three terms are related to $T_1$.
We observe that the smaller the $d_{01}$, the larger the value of $\delta_0$.
And the bigger the $p_{1}^{'}$, the larger the value of $\delta_0$.
So $d_{01}$ has a negative correlation with $\delta_0$ while $p_{1}^{'}$ has a positive correlation.
$\frac{p_1^{'}}{d_{01}}$ is a combination of two terms to increase the number of features.
We also tried a few other features, but we decided to use only these three features for a close by transmitter as a balance of model accuracy and model complexity.

Equation~\ref{equ:twotxpower} is for the case of one close by transmitter, we then extend the equation to handle multiple close by transmitters in the following Equation~\ref{equ:multitxpower},
\begin{equation}
  \delta_0 =  \theta_0 \cdot p_0^{'} + \sum_{i=1}^{m} ( \theta_{(i,1)} \cdot d_{0i} + \theta_{(i,2)} \cdot p_i^{'} + \theta_{(i,3)} \cdot \frac{p_i^{'}}{d_{0i}})
  \label{equ:multitxpower}
\end{equation}
where $m$ is the number of close by transmitters for $T_0$, the transmitter of interest, $d_{0i}$ is the distance between $T_0$ and close by $T_i$, and $p_i^{'}$ is the uncorrected power predicted by \power. 
For the $i$th close by transmitter, we introduce three terms $ d_{0i},  p_i^{'}, \frac{p_i^{'}}{d_{0i}}$, and assign three coefficients $ \theta_{(i,1)}, \theta_{(i,2)}, \theta_{(i,3)}$ to the three terms respectively.
So for $m$ close by transmitters, there are $1 + 3m$ number of terms in the Equation~\ref{equ:multitxpower}.

After modeling $\delta_0$, in Equation~\ref{equ:correct}, we ``correct'' $p_0^{'}$ by subtracting $\delta_0$ from $p_0^{'}$ to get more an accurate estimation of the power of transmitter $T_0$.
\begin{equation}
    p_{0}^{correct} = p_{0}^{'} - \delta_0
    \label{equ:correct}
\end{equation}

\softpara{Estimating the parameter $\theta$}.
Equation~\ref{equ:multitxpower} is essentially a linear model and we can train it by using either linear, ridge, or LASSO regression models~\cite{scikit-learn}.
We perform experiments using ridge regression (alpha=0.01). 
We set a distance threshold for a neighbor transmitter to be classified as a close by transmitter. 
Note that the transmitters will have a different number of close by transmitters. 
So, let us denote $M$ as the maximum number of close by transmitters we see in the dataset.
When training the linear model in Equation~\ref{equ:multitxpower}, we train a model that assumes a maximum $M$ number of close by transmitters, i.e., the linear model has $1+3M$ terms.
The $3M$ terms are organized in a group of three (i.e., three features) and the groups are sorted by distance in an ascending order.
Then, for a transmitter with a smaller than $M$ number of close by transmitters, let us say $m$, only the first $1+3m$ terms will have a meaningful value.
And for the rest $3(M-m)$ terms, we set the value to zero, i.e., impute missing value with zero.

\section{\bf Evaluation}
\label{sec:evaluation}

To evaluate the performance of our proposed techniques, we conduct large-scale simulations over
two settings based on two different propagation models. In particular, we consider the log-distance-based
propagation model and the Longley--Rice model obtained from SPLAT!~\cite{splat}. We evaluate various 
algorithms, using multiple performance metrics as described below. 

\para{Performance Metrics.} We use the following metrics 1, 2, and 3 to evaluate the localization methods and use the 4th metric to evaluate the power estimation methods.
\begin{enumerate}
    \item Localization Error ($\lerr$)
    \item Miss rate ($\mr$)
    \item False Alarm rate ($\fr$)
    \item Power Error ($\perr$)
\end{enumerate}
Given a multi-transmitter localization solution, we first compute the $\lerr$ as
the minimum-cost matching in the bi-partite graph over the
ground truth and the solution's locations, where the cost of each edge in the graph is the Euclidean distance between the matched ground truth node location and the solution's node location.
We use a simple greedy algorithm to compute the min-cost matching.
The unmatched nodes are regarded as false alarms or misses. 
We also put an upper threshold on the cost ($\lerr$) of an eligible match. 
E.g., if there are four intruders in reality, but the algorithm predicts six
intruders then it is said to incur zero misses and two false alarms, so the $\mr$ is zero and the $\fr$ is one-third. 
If the algorithm predicts three intruders then it incurs one miss and zero false alarms, so the $\mr$ is one-fourth and the $\fr$ is zero.
In the plots, we stack the miss rate and false alarm rate to reflect the overall performance.

\para{Algorithms Compared.} 
We implement\footnote{Source code at: \url{https://github.com/caitaozhan/deeplearning-localization}.} and compare six algorithms in two stages. In stage one, we compare three versions of our
techniques, viz., \our, \ouryolo, and \ourpeak.  Recall that \our, \ouryolo, and \ourpeak 
use \imgimg in the first step, and \yolocust, original YOLOv3, and \simpeak respectively in the second step. In the first stage of our evaluations, we will show that 
\our outperforms \ouryolo and \ourpeak in almost all performance metrics. 
Thus, in the second stage, we only compare \our with
schemes from three prior works, viz., \splot~\cite{mobicom17-splot}, \deeptx~\cite{icccn20-deeptxfinder},
and \map~\cite{ipsn20-mtl} and show that \our outperforms the prior works.

\para{Training and Testing Dataset.}
We consider an area
of $1km \times 1km$, and use grid cells (pixels) of 
$10 m \times 10 m$, so the grid is $100\times100$. 
The transmitters may be deployed
anywhere within a cell (i.e., their location is in the continuous
domain), while the sensors are deployed at the centers of the grid cells (i.e. their location is in the discrete domain).
For each instance (training or test sample), the said number of sensors and transmitters are deployed in the field randomly. 
For each of the two settings (propagation models described below), we create a 100,000 sample training dataset to train our models and create another 20,000 sample testing dataset to evaluate the trained model. 

We will evaluate the performance of various techniques for
varying number of transmitters/intruders and sensor
density. When we vary a specific parameter, the other parameter
is set to its \emph{default} value; the number of transmitters varies from 1 to 10 and the default value is 5; the sensor density varies from 1\% to 10\% and the default value is 6\% (600  sensors in a $100\times100$ grid).
The two default numbers 5 and 6\% are chosen because they are in the middle of their ranges.
When not mentioned, the default values are used.
The transmitter power varies from 0 to 5 dBm and is randomly picked.
To minimize overfitting, the training dataset and testing dataset have sensors placed at completely different locations.

We train the \our model using the 100,000 sample dataset. To train \deeptx~\cite{icccn20-deeptxfinder}, we partition the 100,000 sample training dataset into ten datasets based on the number of transmitters in the samples which varies from 1 to 10. These ten datasets are used to train the ten ``localization" CNN models in \deeptx, and the full dataset of 100,000 samples is used to train the \deeptx model that determines the number of transmitters. 
For the \map scheme~\cite{ipsn20-mtl}, we assume the availability of all required probability distributions. We note that using a simple cost model (number of samples need to be gathered), 
the overall training cost for \map is an order of magnitude higher than \our and \deeptx. 
Lastly, \splot~\cite{mobicom17-splot} does not require any training.


\para{Two Propagation Models and Settings}.
The sensor readings (i.e. the dataset) are simulated based on a propagation model. 
To demonstrate the generality of our techniques, we consider two propagation models as described below. 

\softpara{Log-Distance Propagation Model and Setting.}
Log-Distance propagation model is a generic model that extends Friis Free space model 
which is used to predict the path loss for a wide range of environments. 
As per this model, the path loss (in dB) between 
two points $x$ and $y$ at a distance $d$ is given by: $PL_d = 10\alpha\log{d} + \mathcal{X},$
where $\alpha$ (we use 3.5) is the path-loss exponent and $\mathcal{X}$ represents the shadowing effect that can be represented by a zero-mean Gaussian distribution with a certain (we use 1) standard deviation. 
Power received (in dBm) at point $y$ due to a transmitter at point $x$ with a 
transmit power of $P_x$ is thus: $P_{x} - PL_d$.
Power received at point $y$ due to multiple sources is assumed to be just an 
aggregate  of the powers (in linear) received from each of the sources.


\softpara{SPLAT! Model and Setting.}
This is a complex model of wireless propagation based on many parameters including locations, terrain data, obstructions, soil conditions, etc.
We use \splat~\cite{splat} to generate path-loss values. \splat is an open-source software implementing the Longley-Rice ~\cite{chamberlin82} Irregular Terrain With Obstruction Model (ITWOM) model.
We consider a random area in Long Island, New York of $1km \times 1km$ large and use the 600 MHz band to generate path losses.


\begin{figure}[h]
	\centering
	\includegraphics[width=0.6\textwidth]{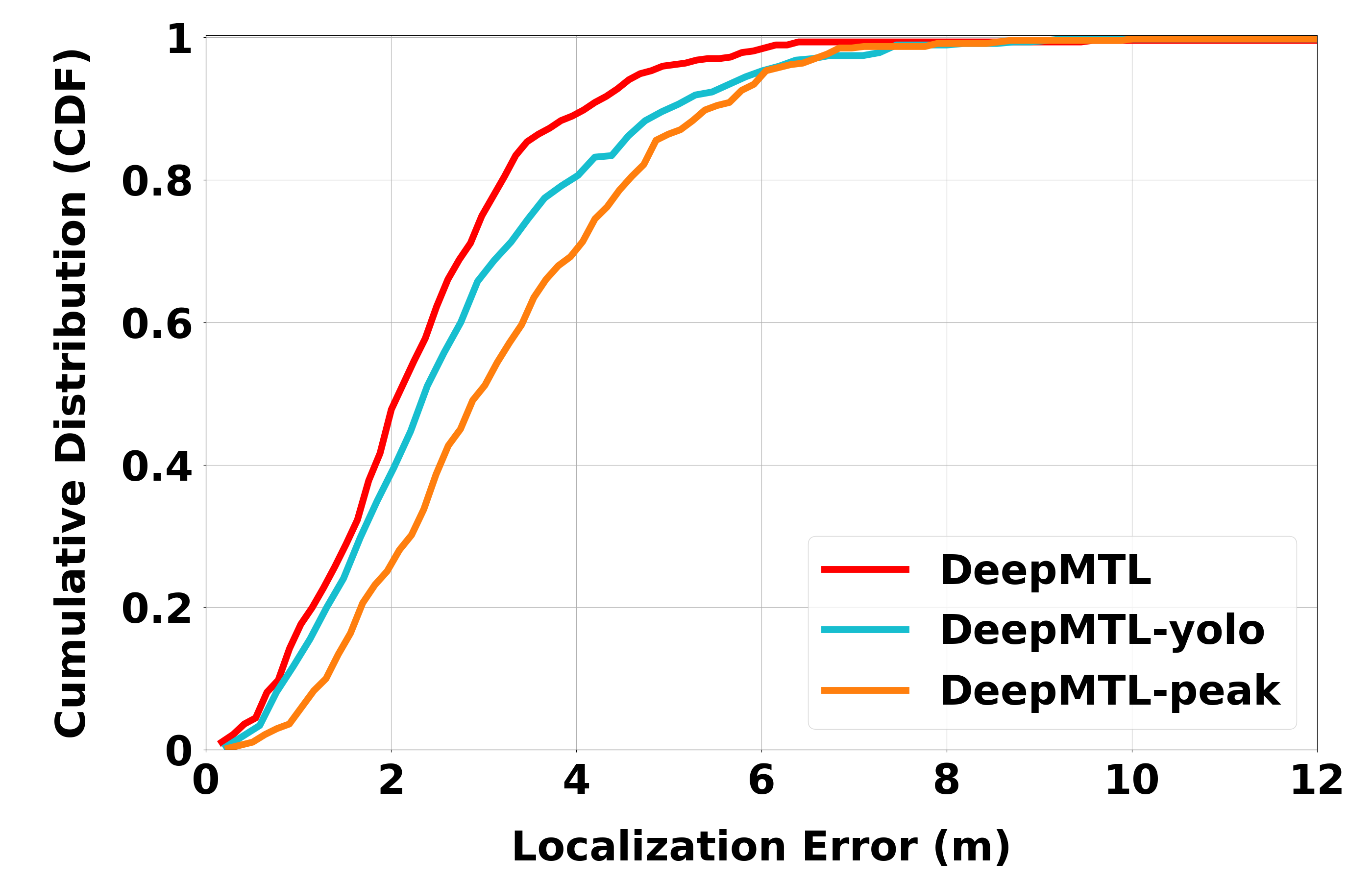}
    \vspace{-0.1in}
	\caption{Cumulative probability of localization error of \our, \ouryolo and \ourpeak, for the special case of single transmitter localization with 6\% sensor density.}
	\label{fig:ours_cdf}
\end{figure}

\begin{figure}[h]
	\centering
	\includegraphics[width=0.75\textwidth]{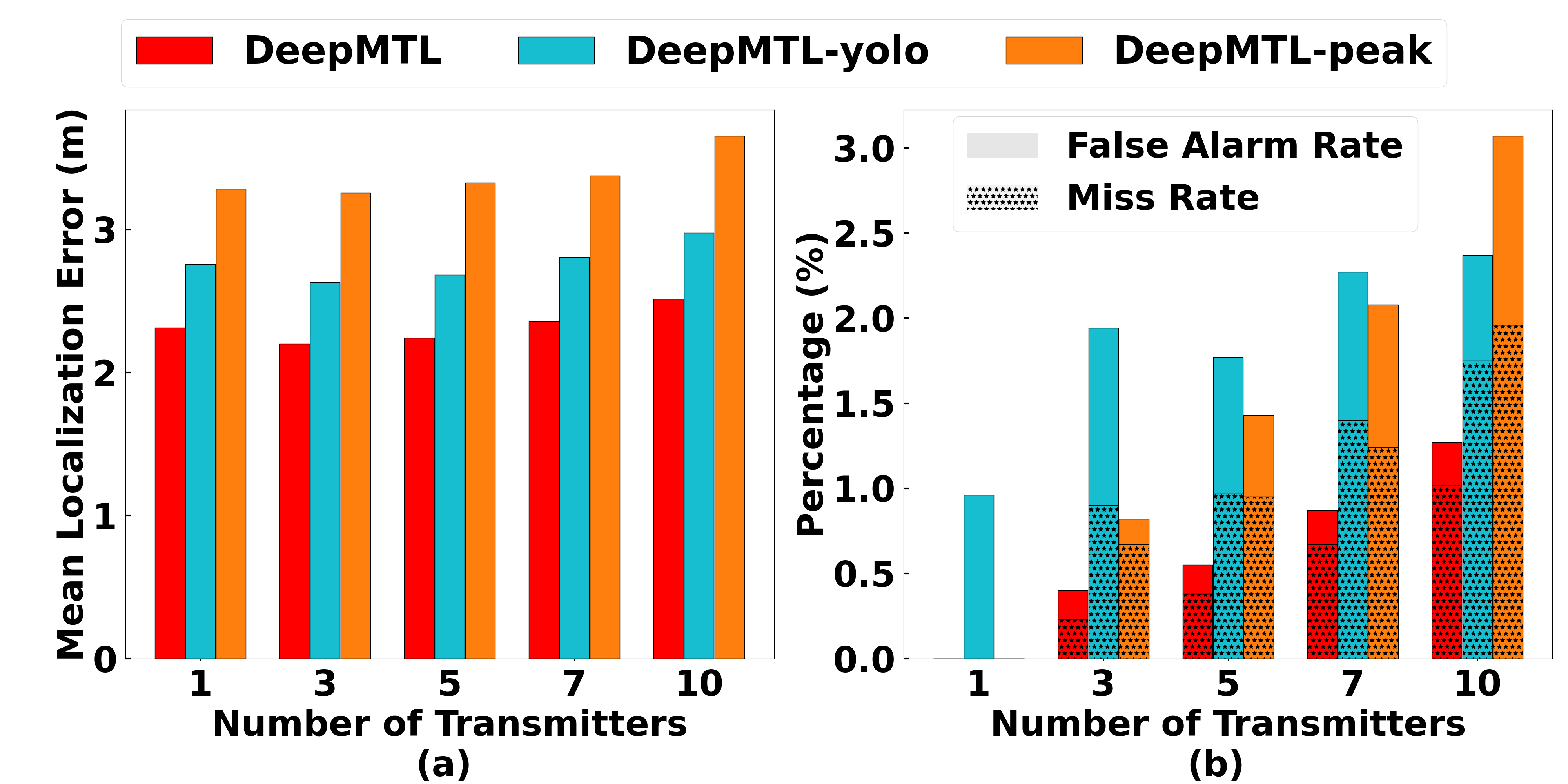}
	\vspace{-0.1in}
	\caption{(a) Localization error and (b) miss and false alarm rates, of \our, \ouryolo and \ourpeak variants for varying number of transmitters in log-distance dataset/propagation model.}
	\label{fig:ours_vary_numintru}
\end{figure}

\begin{figure}[h]
	\centering
	\includegraphics[width=0.75\textwidth]{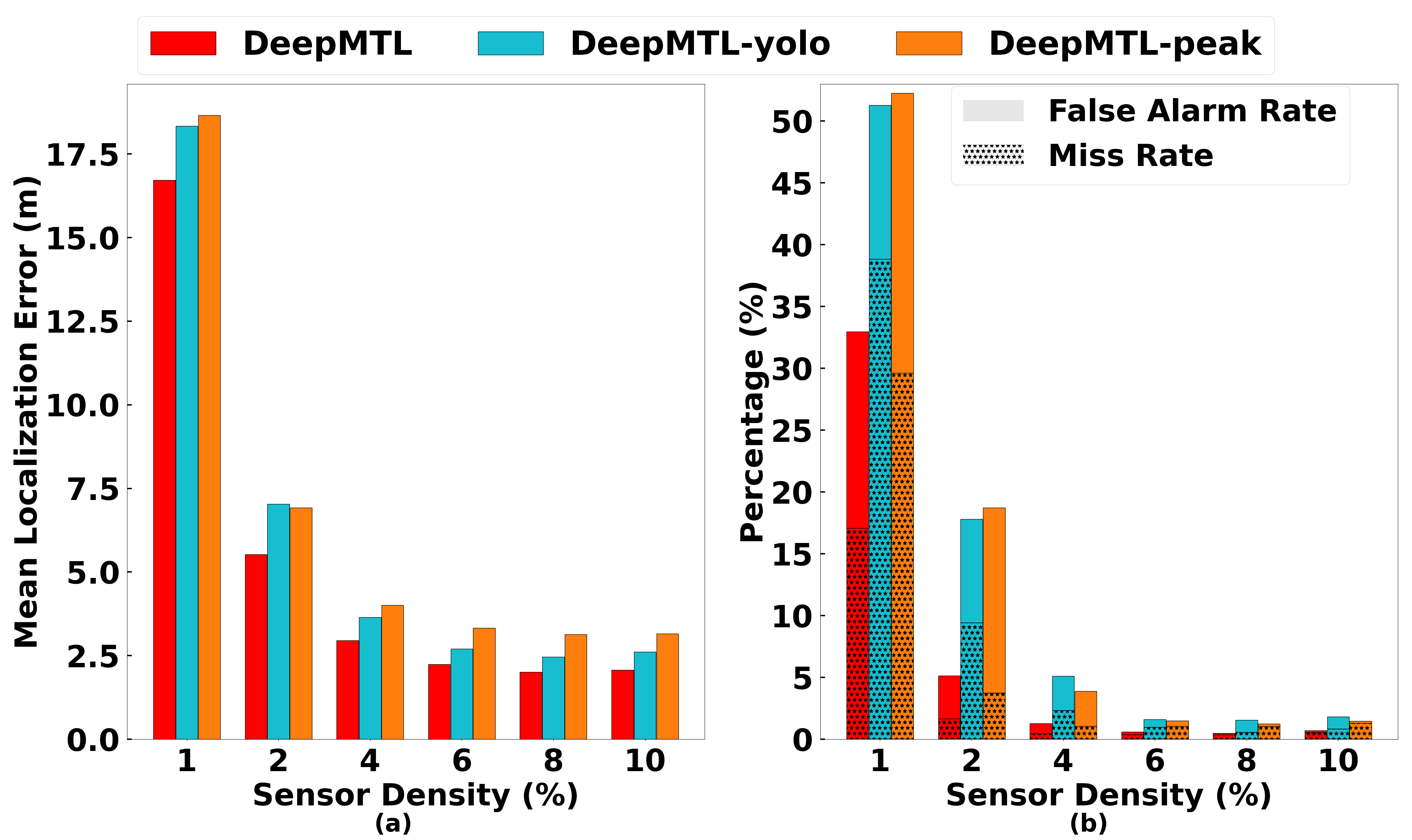}
    \vspace{-0.1in}
	\caption{(a) Localization error and (b) miss and false alarm rates, of \our, \ouryolo and \ourpeak variants for varying sensor density in log-distance dataset/propagation model.}
	\label{fig:ours_vary_sendensity}
\end{figure}

\subsection{\bf \our vs.\ \ouryolo vs.\ \ourpeak}

In this subsection, we compare the three variants of our technique, viz., \our, \ouryolo, and
\ourpeak. For simplicity, we only show plots for the log-distance propagation model setting in this 
subsection (we observed similar performance trends for the Longley-Rice propagation model too). 

\softpara{Performance Results.}
In Fig.~\ref{fig:ours_cdf}, we plot  the cumulative density function (CDF) of the localization
error, for the simple case of a single transmitter.
We observe that \our outperforms the other variants, as it yields a higher cumulative probability for a lower range of errors.
In addition, we evaluate the three variants for varying number of transmitters (Fig.~\ref{fig:ours_vary_numintru}) and sensor density (Fig.~\ref{fig:ours_vary_sendensity}), and evaluate the localization error as well as the false alarm and miss rates. 
We observe that \our consistently outperforms the other two variants 
across all plots and performance metrics. As expected, the performance of all algorithms 
degrades with an increase in the number of transmitters (in terms of false alarms and miss rates) or with a decrease in sensor density. 
In general, the localization error of \our is around 15-30\% lower than the other variants.
Impressively, the total cardinality error (i.e., false alarms plus miss rates) is fewer than 1\% for the \our technique, when the sensor density is 6\% or above.

When the sensor density is as low as 1\%, the performance of all methods significantly decreases.
Because when the sensor density is 1\% or lower, the input image will be very sparse and contain only a few pixels.
\our's first part \imgimg has a receptive field of $17\times17$.
This area will contain an average of less than three sensors when the sensor density is 1\% ($17\times17\times0.01=2.89$).
This number is considered too low and note that 2.89 sensors are not enough for the trilateration localization method, which needs three sensors.
Our CNN models need to function well with enough pixels that contain useful information. 
So we suggest the sensor density to be at least 2\% to achieve reasonable results.

\begin{table}[h]
	\centering
	\caption{Compare Localization Running Time (s) for 1 to 10 Number of Intruders}
	\vspace{-0.1in}
	\begin{tabular}{c c c c c c c}
		\hline\hline
		\small{Intru.} & \small{\ourpeak}  & \small{\ouryolo} & \small{\our} & \small{\map} & \small{\splot} & \small{\deeptx} \\
		\hline
		1 & 0.0013 & 0.0180 & 0.0180 & 8.78 & 1.53 & 0.0015 \\ 
		3 & 0.0014 & 0.0183 &  0.0186 & 15.1 & 1.79 & 0.0016 \\
		5 & 0.0016 & 0.0192 &  0.0189 & 19.3 & 2.06 & 0.0017\\
		7 & 0.0018 & 0.0196 &  0.0194 & 24.1 & 2.32 & 0.0019 \\
		10 & 0.0023 & 0.0205 & 0.0206 & 28.5 & 2.72 & 0.0022 \\
		\hline
	\end{tabular}
	\label{table:running-time}	
\end{table}

\softpara{Running Time Comparison.} For the running time comparison of the variants, see Table \ref{table:running-time}. 
Our hardware is an Intel i7-8700 CPU and an Nvidia RTX 2070 GPU. 
We observe that, as expected, \our and \ouryolo which use a sophisticated object-detection method do incur higher latency (around 20 milliseconds) than \ourpeak (around two milliseconds). As our key
performance criteria is accuracy and the run time of \our is still quite low, we choose \our for comparison with the prior works in ~\S\ref{subsec:vs_prior}.

\softpara{Localizing Transmitters Close By.}
Localizing two or more transmitters close by is a hard part of the \mtl problem.
Fig.~\ref{fig:peaks}(c) and (d) gives an example of when an advanced object detection algorithm will work while a simple local maximal peak detection might not.
Fig.~\ref{fig:peaks}(c) and (d) shows \our can successfully localize two transmitters as close as three pixels apart.
When a pixel represents a $10m \times 10m$ area, then it is 30 meters apart.
If a pixel represents a smaller area, such as $1m \times 1m$, it has the potential to localize two transmitters as close as three meters apart.

\softpara{Two YOLO Thresholds.} YOLO has two important thresholds to tune that can affect the miss rate and false alarm rate. 
One is the confidence threshold (\conf) and the other is the non-maximum suppression threshold (\nms).
An object will be recognized as a peak only if its confidence level is larger than \conf.
If two recognized peaks' bounding boxes have a large overlap, and their intersection of union is higher than \nms, then the two peaks will be considered as one peak. 
The peak with a higher confidence level keeps while the other peak with a lower confidence level discards.
A higher \conf will bring a lower false alarm rate but a higher miss rate, and a higher \nms will bring a lower miss rate but a higher false alarm rate.
We pick \conf= 0.8 and \nms= 0.5 for \our as we observe these values bring a good balance between false alarm rate and miss rate.
In particular, a high \conf of 0.8  precludes ``fake peaks" at locations with no transmitters.
Also, a low \nms weakens \our's ability to localize two close by transmitters, while a high $nms$ yields
a high false alarm rate (by incorrectly interpreting a single transmitter as multiple close by transmitters); thus, we chose \nms of 0.5.


\subsection{\bf \our vs.\ Prior Works}
\label{subsec:vs_prior}

\begin{figure}[t]
	\centering
	\includegraphics[width=0.75\textwidth]{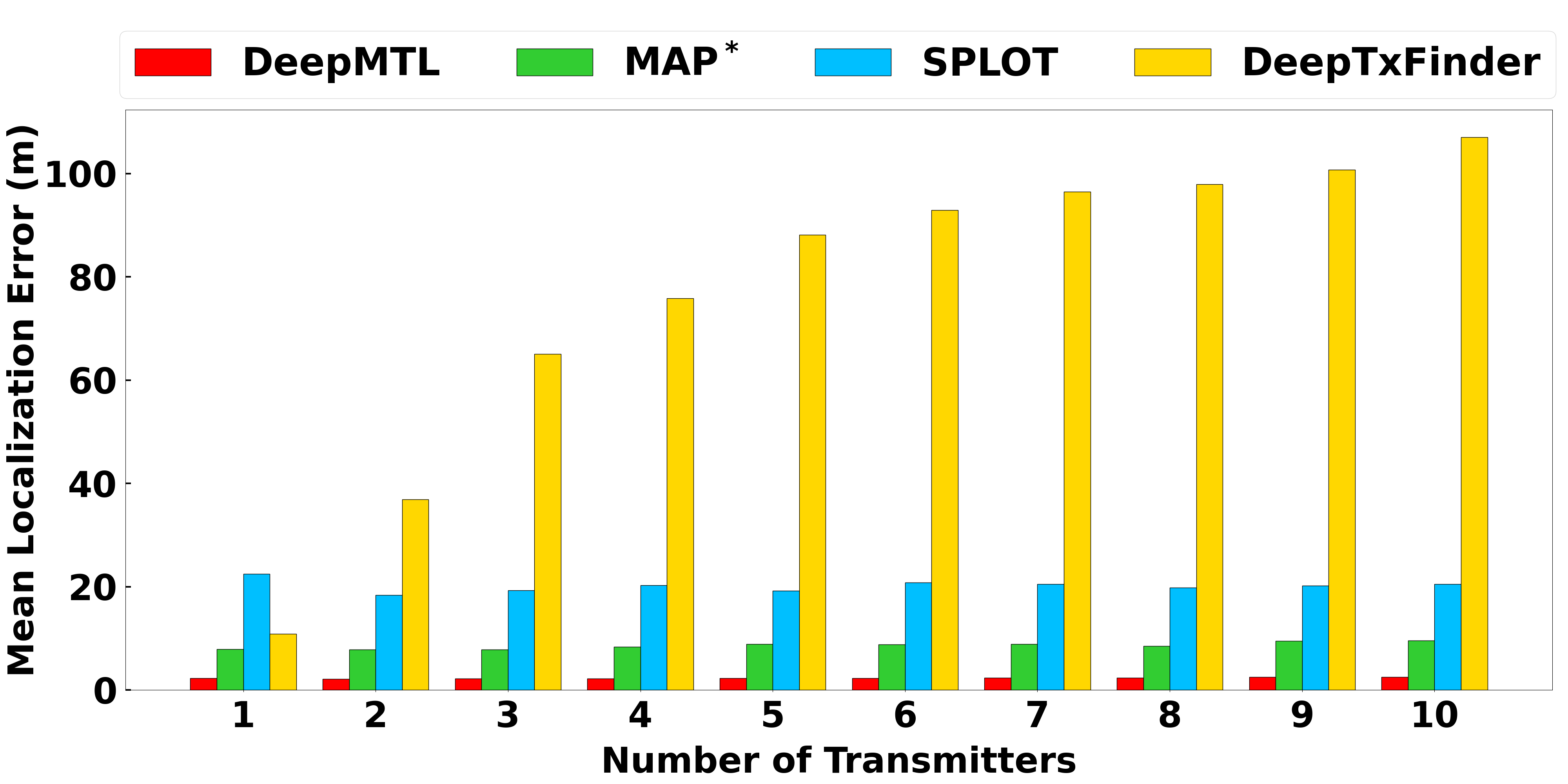}
    \vspace{-0.1in}
	\caption{Localization error of \our, \map, \splot, and \deeptx for varying number of transmitters in the log-distance dataset.}
	\label{fig:logdist-error-vary_numintru}
\end{figure}

\begin{figure}[t]
	\centering
	\includegraphics[width=0.75\textwidth]{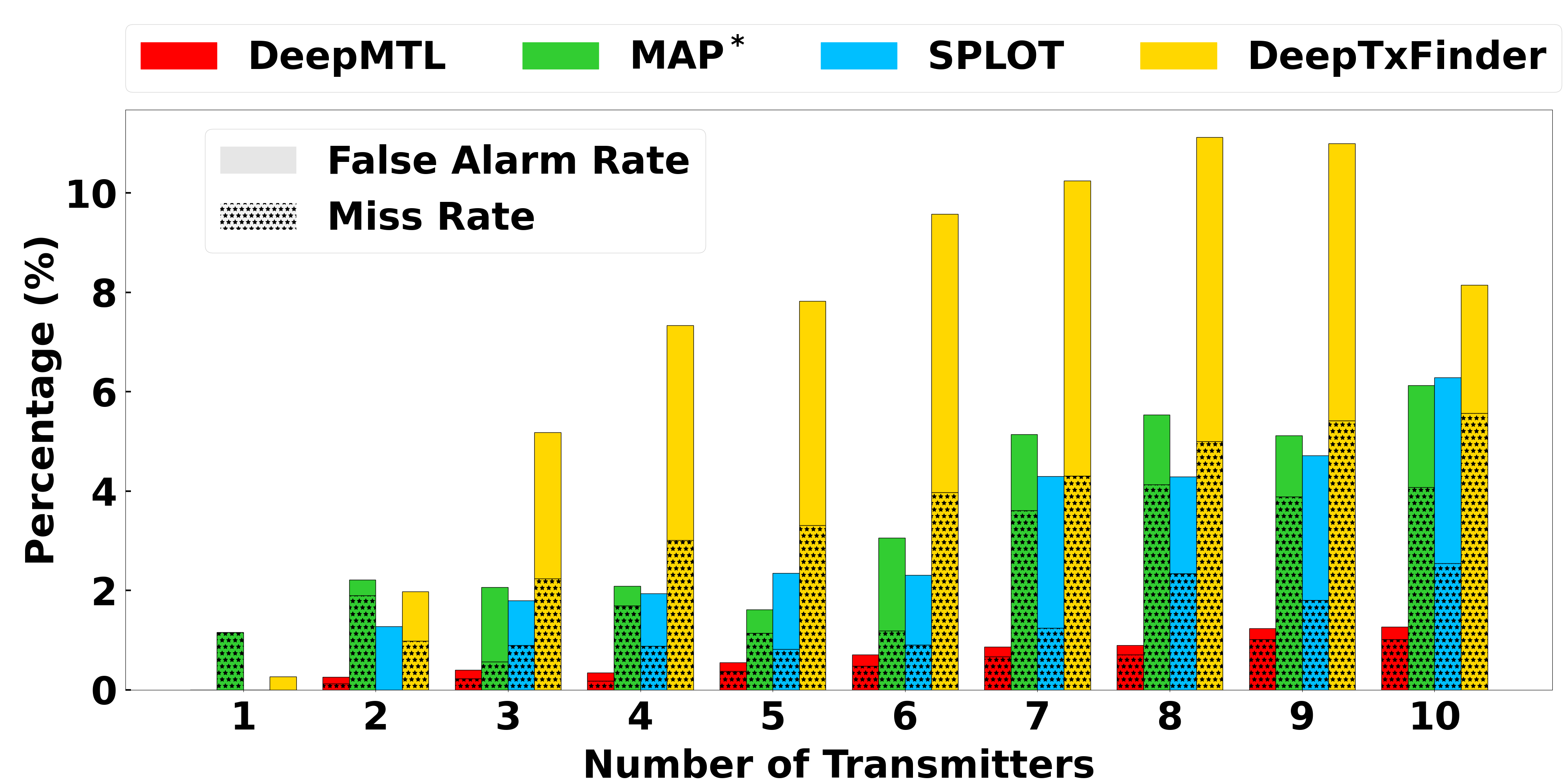}
    \vspace{-0.1in}
	\caption{Miss and false alarm rates of \our, \map, \splot, and \deeptx for varying number of transmitters in the log-distance dataset.}
	\label{fig:logdist-missfalse-vary-numintru}
\end{figure}

\begin{figure}[h]
	\centering
	\includegraphics[width=0.75\textwidth]{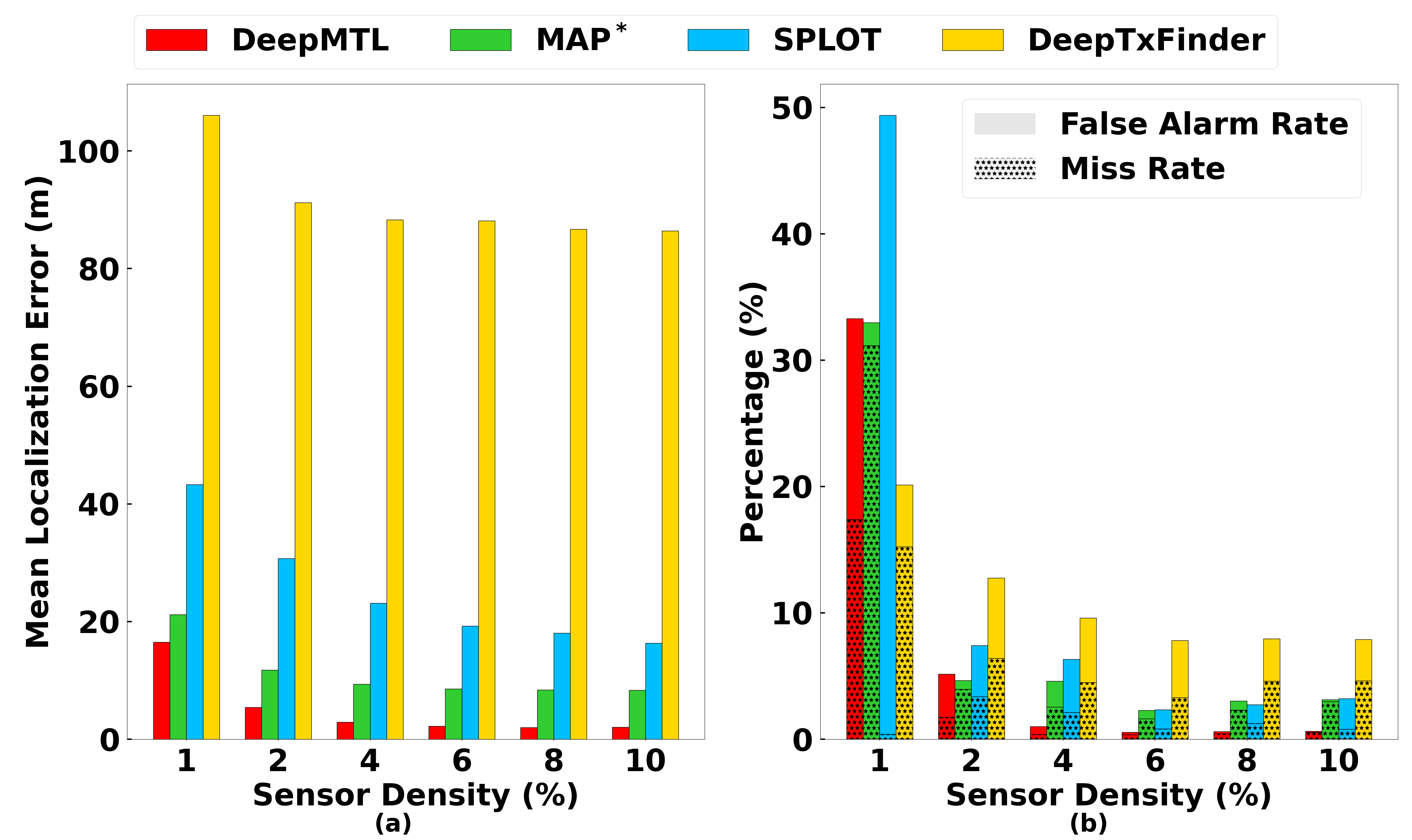}
    \vspace{-0.1in}
	\caption{(a) Localization error, and (b) miss and false alarm rates, of \our, \map, \splot, and \deeptx for varying sensor densities in the log-distance dataset.}
	\label{fig:logdist-error_missfalse-vary-sendensity}
\end{figure}


\begin{figure}[h]
	\centering
	\includegraphics[width=0.75\textwidth]{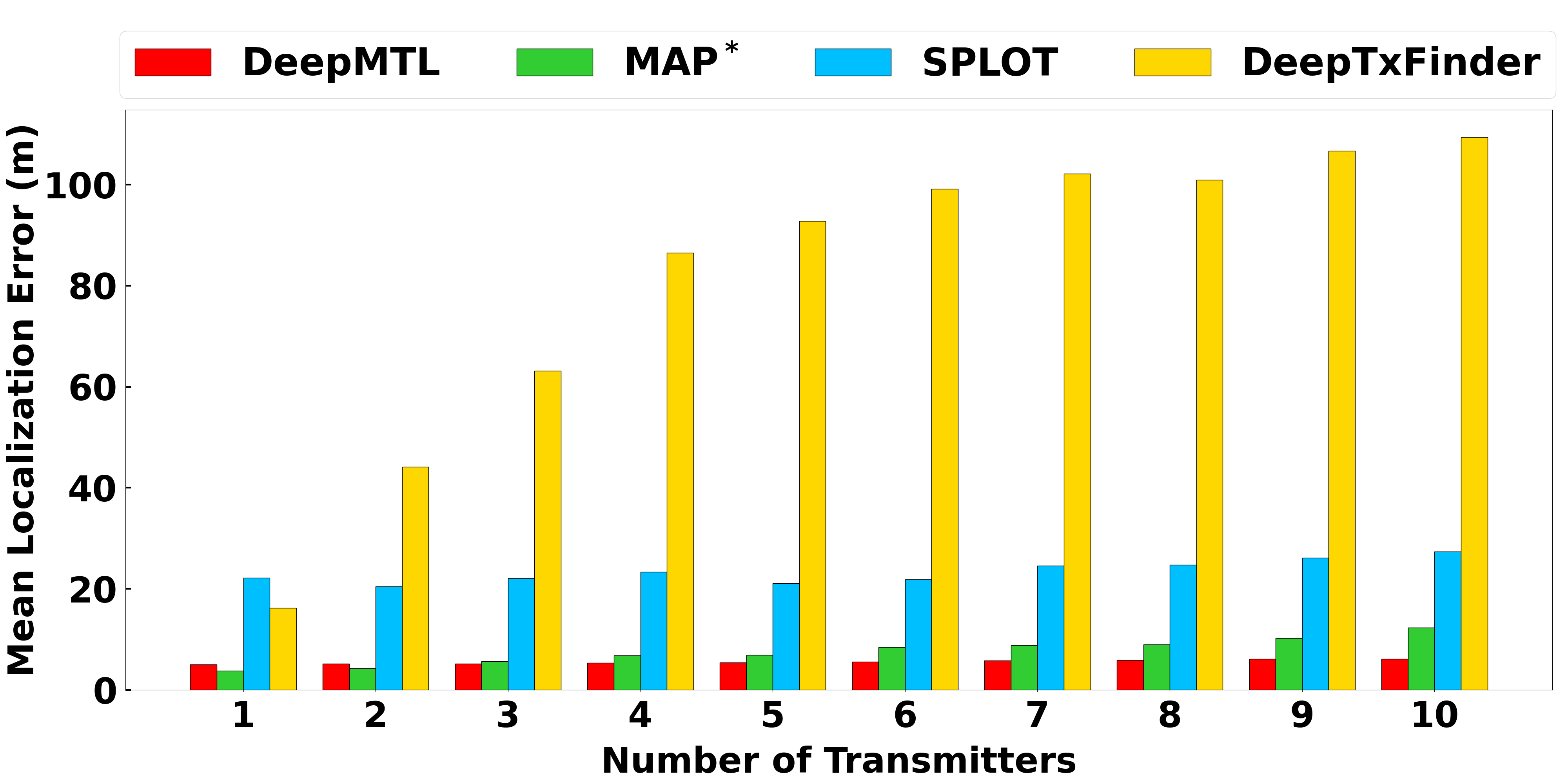}
    \vspace{-0.1in}
	\caption{Localization error of \our, \map, \deeptx and \splot for varying number of transmitters in the SPLAT! Dataset. }
	\label{fig:splat-error-vary_numintru}
\end{figure}

\begin{figure}[h]
	\centering
	\includegraphics[width=0.75\textwidth]{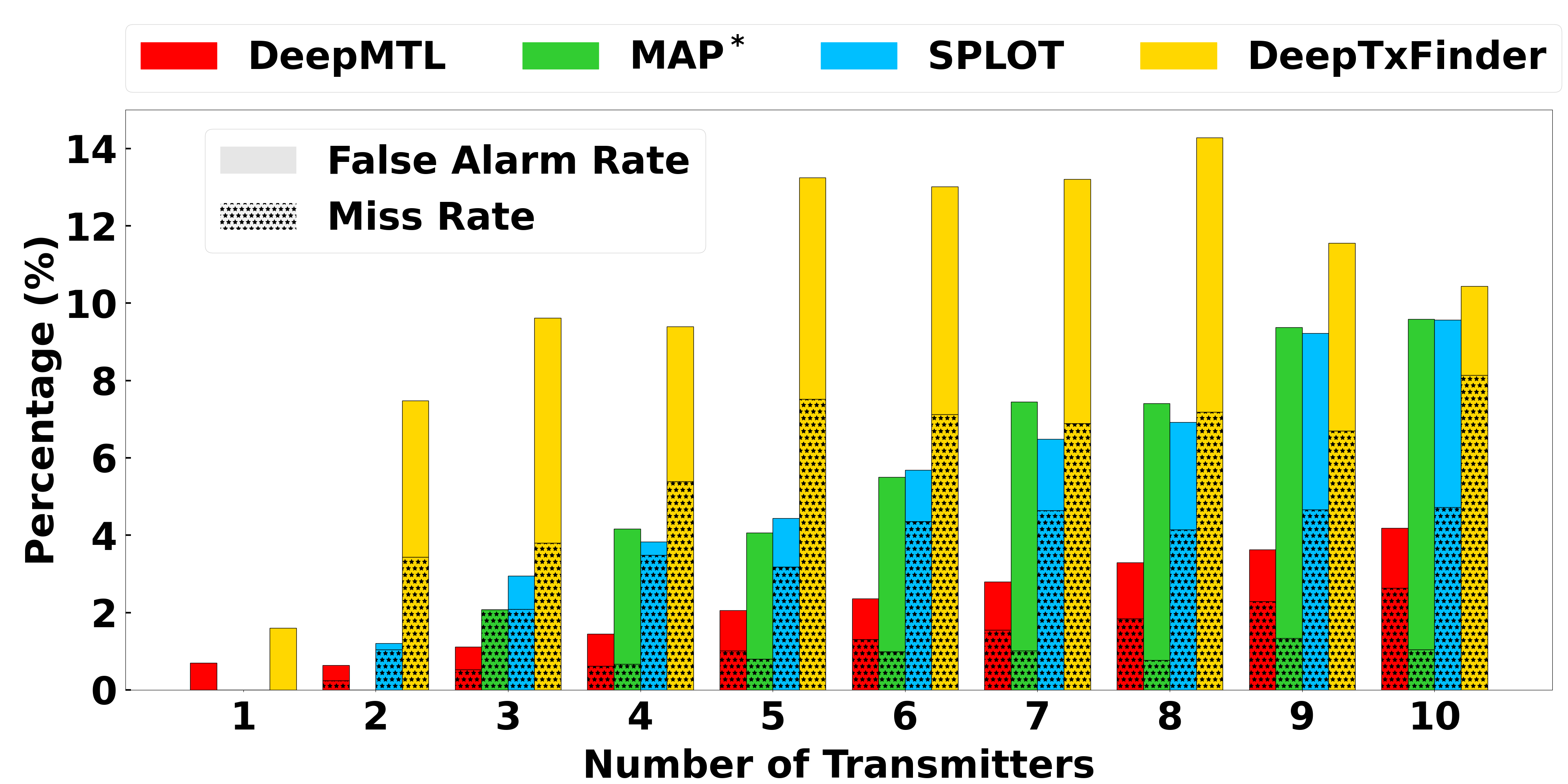}
    \vspace{-0.1in}
	\caption{Miss and false alarm rates of \our, \map, \splot, and \deeptx for varying number of transmitters in the SPLAT! Dataset.}
	\label{fig:splat-missfalse-vary-numintru}
\end{figure}

\begin{figure}[h]
	\centering
	\includegraphics[width=0.75\textwidth]{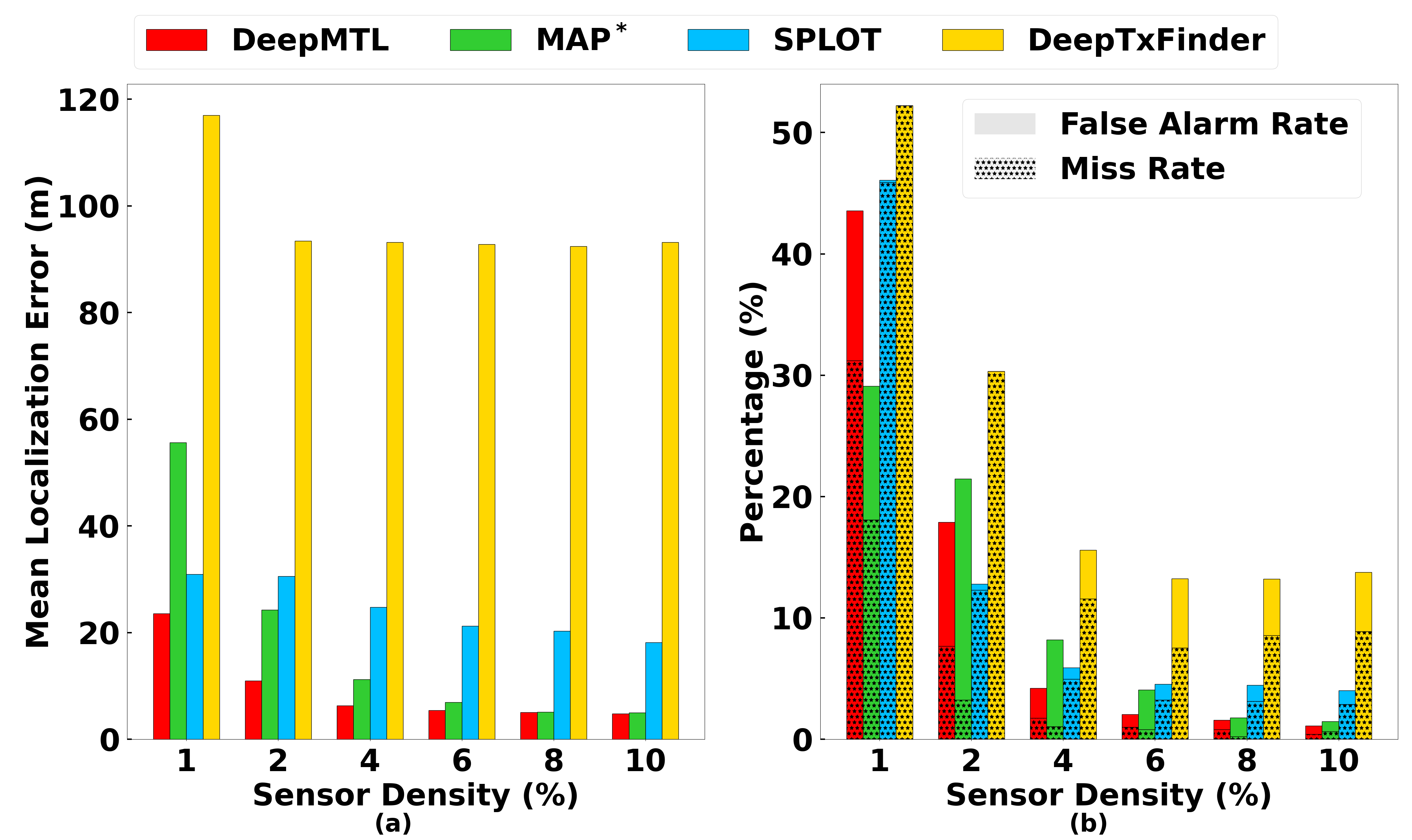}
    \vspace{-0.1in}
	\caption{(a) Localization error, and (b) miss and false alarm rates, of \our, \map, \splot, and \deeptx for varying sensor densities in the SPLAT! Dataset.}
	\label{fig:splat-error-vary-sendensity}
\end{figure}

In this subsection, we compare \our with \splot, \map, \deeptx in both log-distance (Fig.~\ref{fig:logdist-error-vary_numintru}, \ref{fig:logdist-missfalse-vary-numintru}, \ref{fig:logdist-error_missfalse-vary-sendensity}) and SPLAT (Fig.~\ref{fig:splat-error-vary_numintru}, \ref{fig:splat-missfalse-vary-numintru}, \ref{fig:splat-error-vary-sendensity}) propagation models and thus, datasets.
We observe similar performance trends for both datasets, i.e., 
\our significantly outperforms the other approaches by a large margin (in many cases, by more
than 50\% in localization errors, false alarms, and miss rates). For all techniques, as expected,
the performance is generally worse in the SPLAT dataset compared to the log-distance dataset.

\softpara{Varying Number of Transmitters.}
Fig.~\ref{fig:logdist-error-vary_numintru} and Fig.~\ref{fig:splat-error-vary_numintru} show the localization error with varying number of transmitters, in the two datasets. We see that \our has a mean localization error of only 2 to 2.5 meters (roughly, one-fourth of the side length of a pixel/grid cell) in the log-distance dataset and about 5 to 6 meters in the SPLAT dataset. In comparison, the
localization errors of \map, \splot, \deeptx are two to three times, eight to nine times, and 
few tens of times respectively more than that of \our. 
Fig.~\ref{fig:logdist-missfalse-vary-numintru} and Fig.~\ref{fig:splat-missfalse-vary-numintru} show the miss and false alarm rates with varying number of transmitters in the two datasets.
We observe that \our's summation of miss and false alarm rate is only 1\% even at 
ten transmitters in the log-distance dataset, and about 4\% for the case of \splat dataset. 
In comparison, the summation of miss and false alarm rates for other schemes is at least 6\% and
10\% respectively for the two datasets, when there are ten transmitters.

\softpara{Varying Sensor Density.}
Fig.~\ref{fig:logdist-error_missfalse-vary-sendensity} and Fig.~\ref{fig:splat-error-vary-sendensity} plot the performance of various algorithms for varying sensor density in the two datasets. For very low sensor density of 1\%, all algorithms perform 
badly (in comparison with higher sensor densities), but \our still performs the best except that \map performs best at 1\% in terms of false alarm rate and miss rate.
For higher sensor densities, we observe a similar performance trend as above---i.e., \our easily outperforms the other schemes by a large margin.
For the SPLAT! dataset at the 6\% sensor density, the summation of false alarm rate and miss rate is 2\%, which is higher than the 1\% summation for the log-distance dataset.

\softpara{Running Times.}
The run time of \our (in tens of milliseconds) is orders of magnitude faster than \map and \splot (both in seconds). 
See Table \ref{table:running-time}. 
The \our run time is an order of magnitude slower than \deeptx (in a few milliseconds), due to the deep \yolocust taking up over 90\% of the run time.

\softpara{Summary and Analysis}. 
In summary, our approach significantly outperforms the other approaches in all the accuracy performance metrics, as well as in terms of latency. 
In particular, our approach also significantly outperforms the other CNN-based approach \deeptx. 
The main reason for \deeptx's inferior performance is its inability to accurately predict the number of TXs---which 
forms a fundamental component of their technique. In contrast, \our can circumvent explicit pre-prediction
of number of transmitters by using a well-developed object-detection technique which works well for multiple objects 
especially in our context of simple objects.

\subsection{\bf Transfer Learning}
\label{subsec:transfer_learning}

\begin{figure*}[t!]
    \centering
    \begin{subfigure}[t]{0.5\textwidth}
        \centering
        \includegraphics[width=\textwidth]{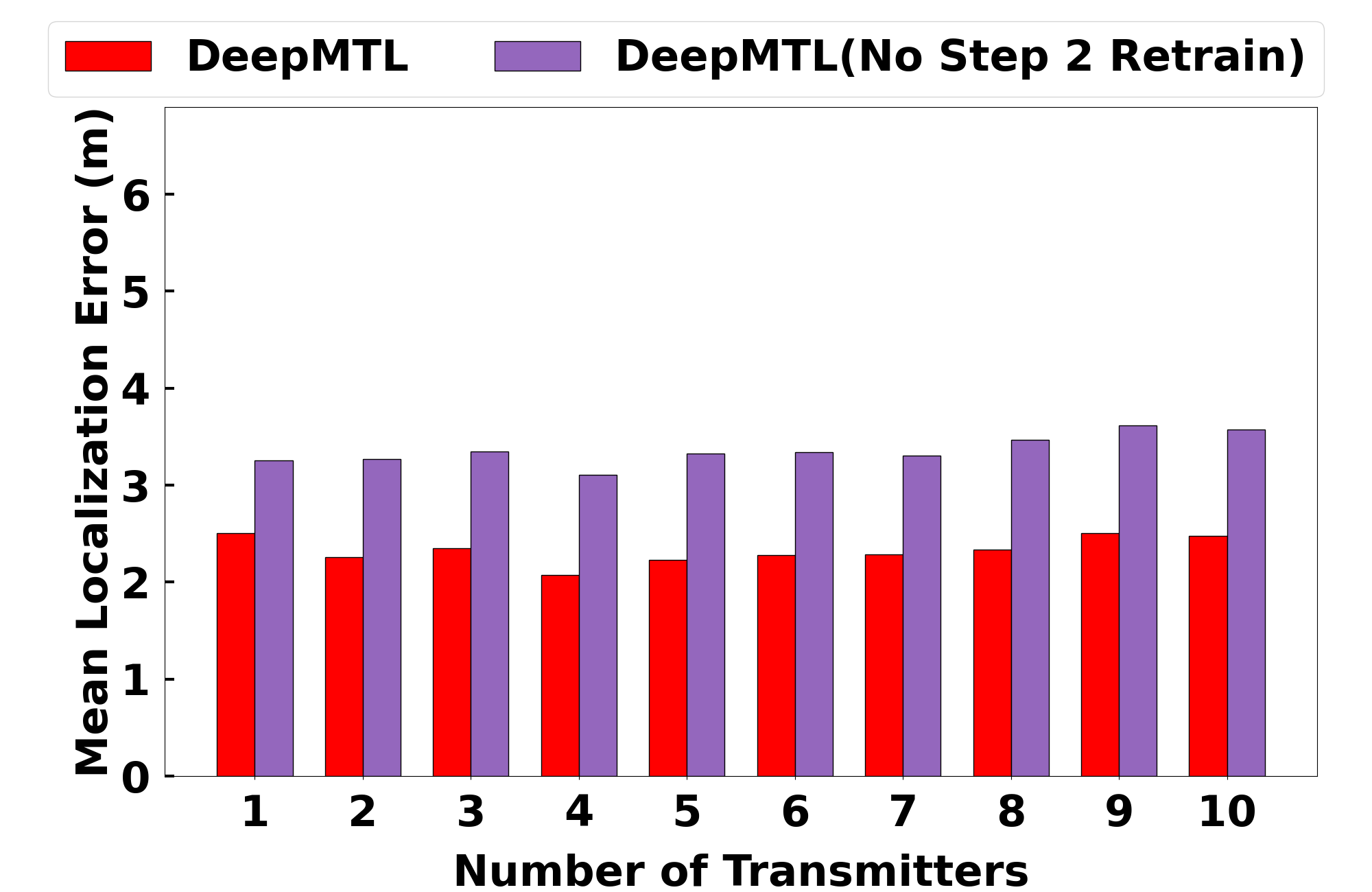}
        \caption{First step trained in log-distance data, while second step trained in SPLAT! data. Tested on the log-distance data.}
        \label{fig:notrain-logdist}
    \end{subfigure}%
    ~ 
    \begin{subfigure}[t]{0.5\textwidth}
        \centering
        \includegraphics[width=\textwidth]{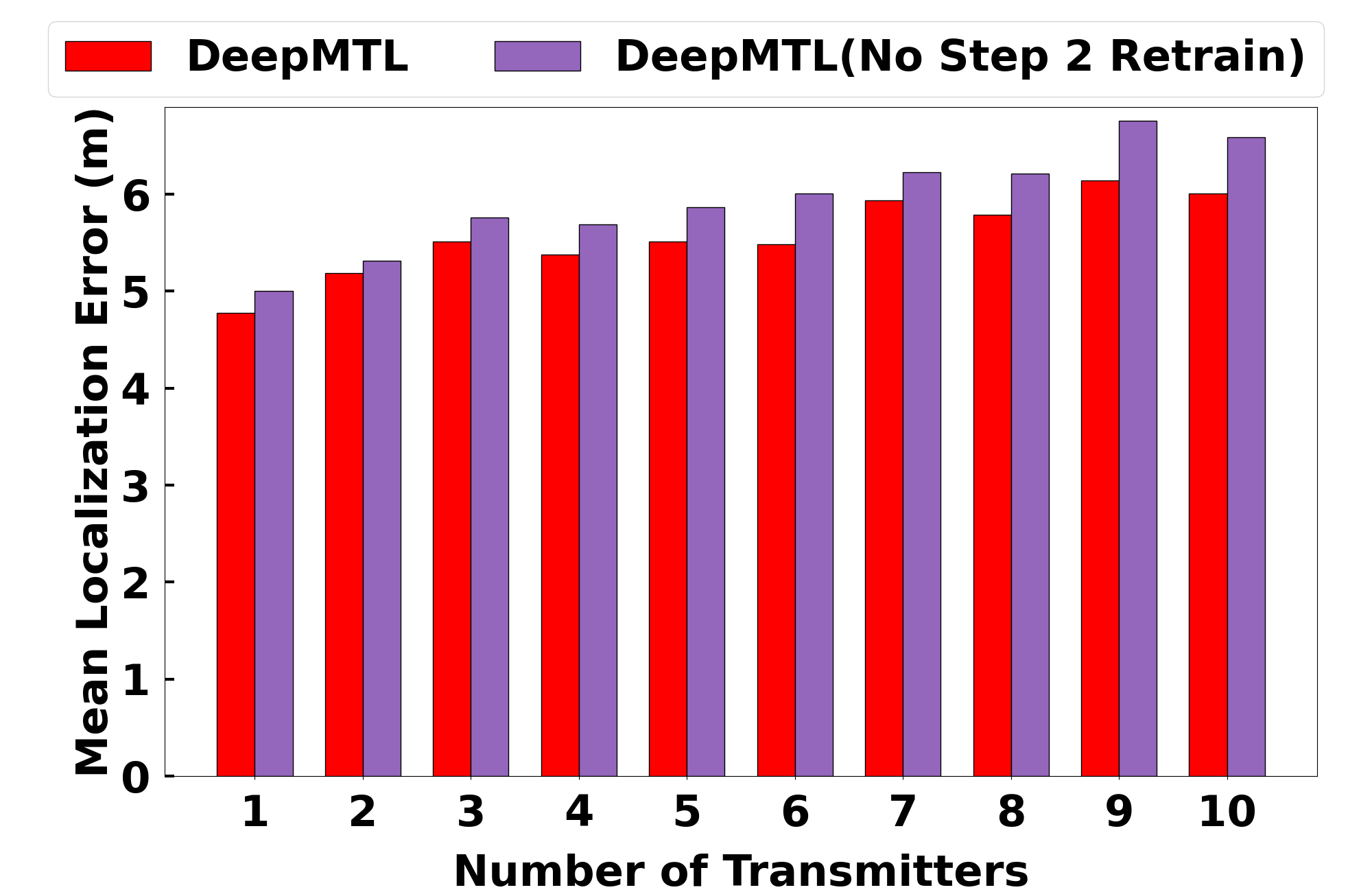}
        \caption{First step trained in SPLAT! data, while second step trained in log-distance! data. Tested on the SPLAT! data.}
        \label{fig:notrain-splat}
    \end{subfigure}
    \caption{Localization error for varying number of transmitters when the first and second step of \our are trained on different training dataset.}
    \label{fig:notrain}
\end{figure*}

\begin{figure*}[t!]
    \centering
    \begin{subfigure}[t]{0.5\textwidth}
        \centering
        \includegraphics[width=\textwidth]{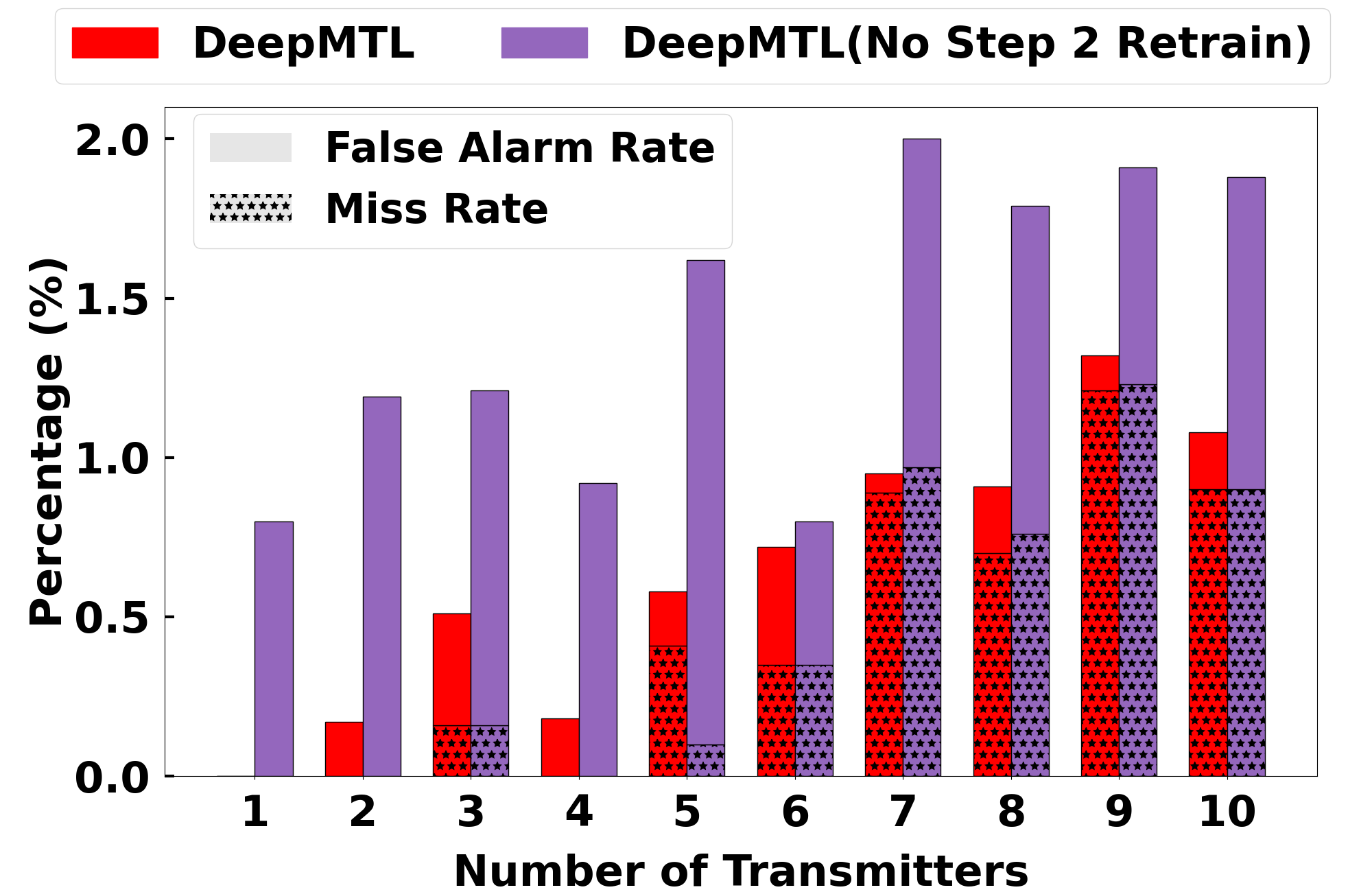}
        \caption{First step trained in log-distance data, while second step trained in SPLAT! data. Tested on the log-distance data.}
        \label{fig:notrain-logdist-missfalse}
    \end{subfigure}%
    ~ 
    \begin{subfigure}[t]{0.5\textwidth}
        \centering
        \includegraphics[width=\textwidth]{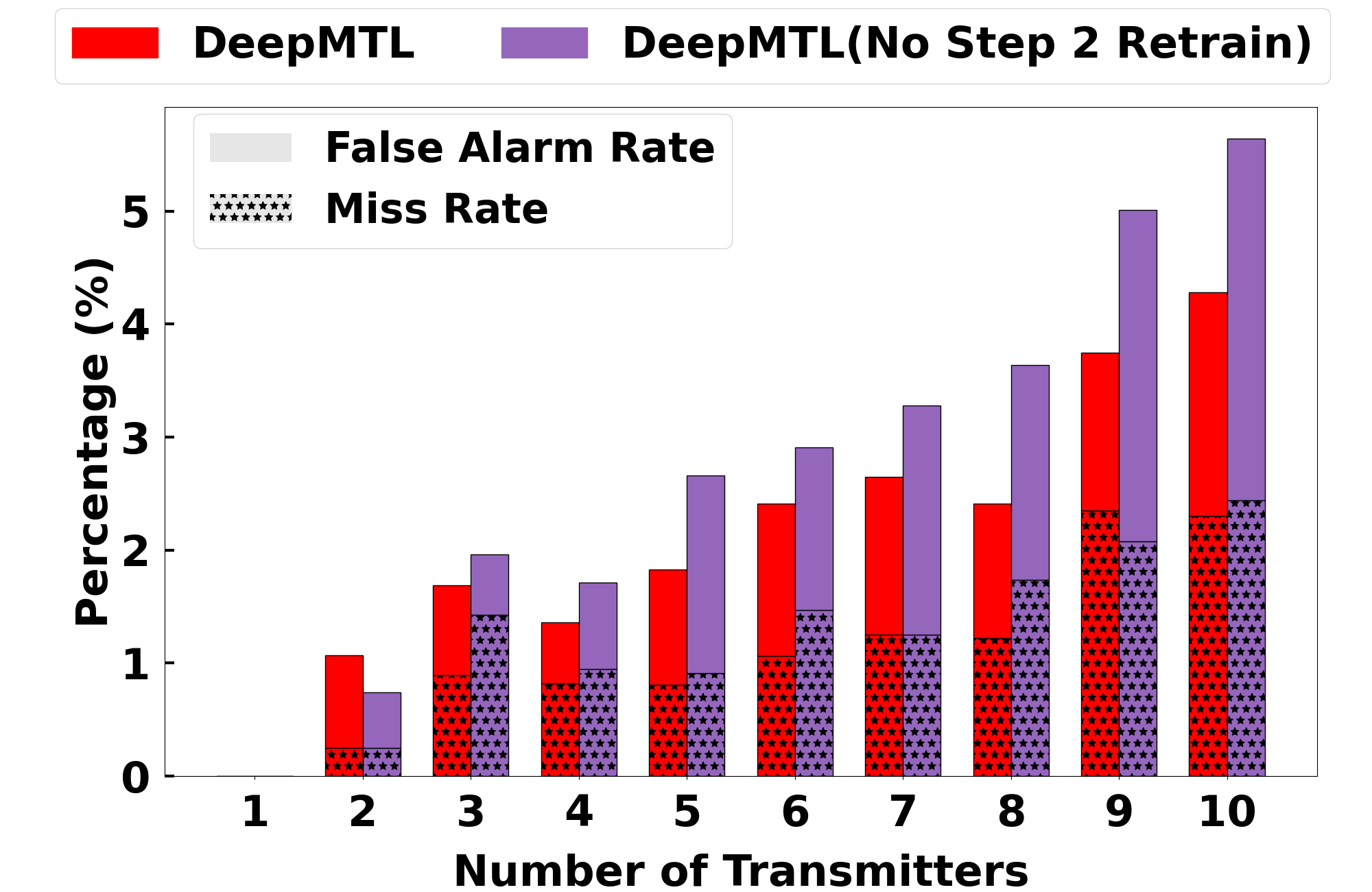}
        \caption{First step trained in SPLAT! data, while second step trained in log-distance! data. Tested on the SPLAT! data.}
        \label{fig:notrain-splat-missfalse}
    \end{subfigure}
    \caption{The miss rate and false alarm rate for varying number of transmitters when the first and second step of \our are trained on different training dataset.}
    \label{fig:notrain-missfalse}
\end{figure*}

We demonstrate transfer learning (generalizability) by showing that the second step in \our does not need to be retrained for different radio frequency propagation models and terrains.
In the previous experiments, the two steps of \our are both trained in the same setting, either log-distance or SPLAT!.
We do the following two combinations to show that the second step does not need to retrain:
\begin{enumerate}
    \item The first step is trained in the log-distance setting and the second step is trained in the SPLAT! setting. Tested on the log-distance data.
    \item The first step is trained in the SPLAT! setting and the second step is trained in the log-distance setting. Tested on the SPLAT! data.
\end{enumerate}
In both combinations, the second step \yolocust is trained on a different dataset compared to the first step \imgimg. 
Fig.~\ref{fig:notrain-logdist} shows that the localization error increases one-third in the first combination compared to the case where both the first and second steps are trained on log-distance dataset. 
Fig.~\ref{fig:notrain-splat} shows that the localization error increases only five percent in the second combination compared to the case where both the first and second steps are trained on SPLAT! dataset.
The miss rate and false alarm rate for both combinations also increase minimally, i.e. the summation of miss rate and false alarm rate only increases around 1\% in absolute value. See Fig.~\ref{fig:notrain-missfalse}.
This implies that the second step of \our, \yolocust, is general and does not need to retrain for different radio frequency propagation models and terrains.
This is because the first step \imgimg is translating sensor readings images from different geographical areas to the same Gaussian peaks.
The first step \imgimg still needs to be retrained for different situations to translate the sensor readings to the peaks.

\subsection{\bf Localize Intruders in the Presence of Authorized Users}
\label{subsec:authorzedeval}

\begin{figure}[t]
    \centering
    \includegraphics[width=0.75\textwidth]{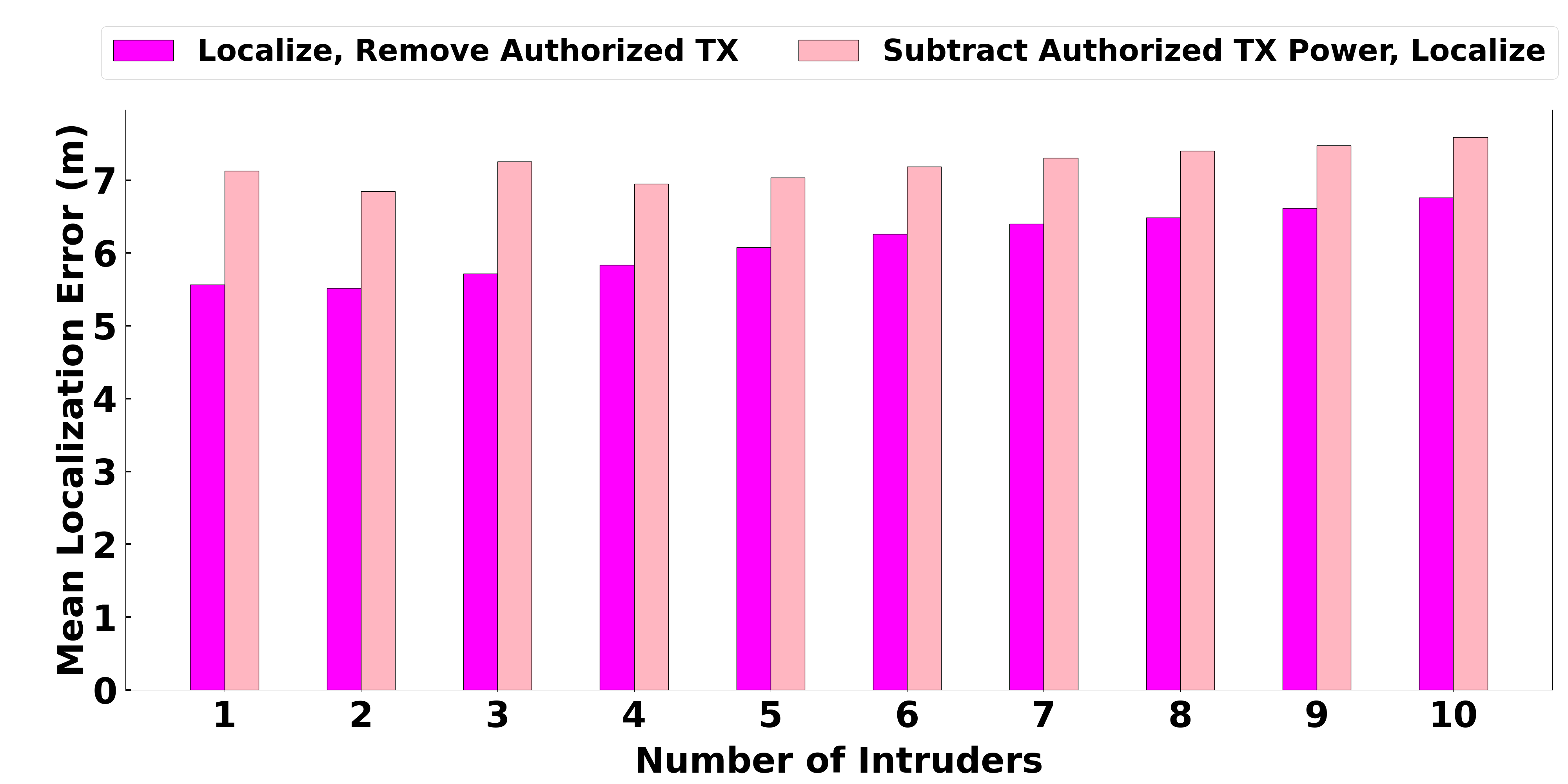}
    \vspace{-0.1in}
    \caption{The localization error of two approaches in the presence of five authorized users with varying number of intruders.}
    \label{fig:authorized_error}
\end{figure}

\begin{figure}[t]
    \centering
    \includegraphics[width=0.75\textwidth]{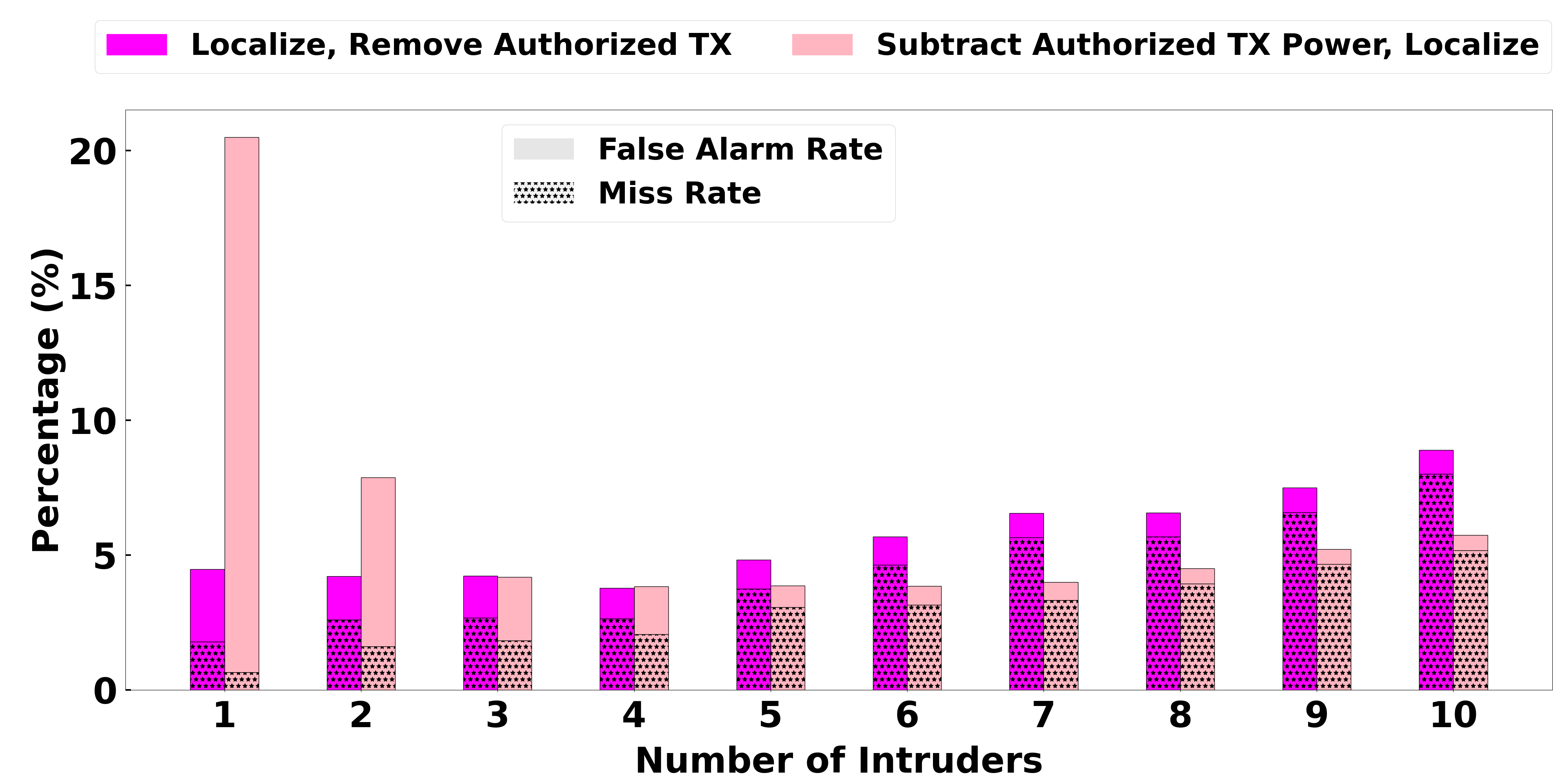}
    \vspace{-0.1in}
    \caption{The miss and false alarm of two localization approaches in the presence of 5 authorized users with varying number of intruders.}
    \label{fig:authorized_missfalse}
\end{figure}

The previous experiment setting is based on the assumption that all transmitters we are localizing are intruders.
Different than the previous setting, here, we put five authorized users and they are spread out in the field, so those five will not interfere with each other.
This is the more general version of the \mtl problem, where there are some authorized users in the background.
Fig.~\ref{fig:authorized_error} shows the localization error of two approaches localizing intruders in the presence of five authorized users with a varying number of intruders.
It is observed that the first approach, localize then remove authorized users, has a ten to twenty percent smaller localization error compared to the second approach, subtract authorized user power then localize.
This is due to the inaccuracy of power subtraction from the \subtract.
Fig.~\ref{fig:authorized_missfalse} shows the miss and false alarm of two approaches localizing intruders in the presence of five authorized users with a varying number of intruders.
It is observed that the second approach, subtract authorized TX power then localize, is having a high false alarm when the number of intruders is three or less.
So for \subtract, subtracting the power of five background authorized users from six transmitters (five out of six transmitters are authorized users, one intruder) is relatively more difficult than subtracting the power of five authorized users from nine users (five out of nine transmitters are authorized users, four intruders). 
Also statistically, getting one false alarm when there are one intruder and five authorized users is 100\% false alarm rate, while getting one false alarm when there are two intruders and five authorized users is only 50\% false alarm rate (the denominator is the number of intruders).
Thus, the false alarm rate for one and two number of intruders looks to differ a lot, but in reality, the false alarm cases do not differ a lot).
When the number of intruders is three or four, the two approaches are comparable. But when the number of intruders is larger than four, the second approach is having a lower miss and false alarm rate.
In summary, the two approaches both have their strengths.
The main advantage for the second approach is that the sum of miss rate and false alarm rate is lower when the number of intruders is large.

\subsection{\bf Power Estimation Evaluation}
\label{subsec:powereval}
In this subsection, we evaluate the transmitter power estimation performance.
In all experiments, the power range is 5 dB.
The power error is presented in absolute value.
A power error of 0.5 dB implies a relative power error of 10\%.
First, we compare the single transmitter power estimation between \map and \power, and then compare the multiple transmitter power estimation between \map, \power with error correction, and \power with error correction.

\begin{figure}
    \centering
    \includegraphics[width=0.75\textwidth]{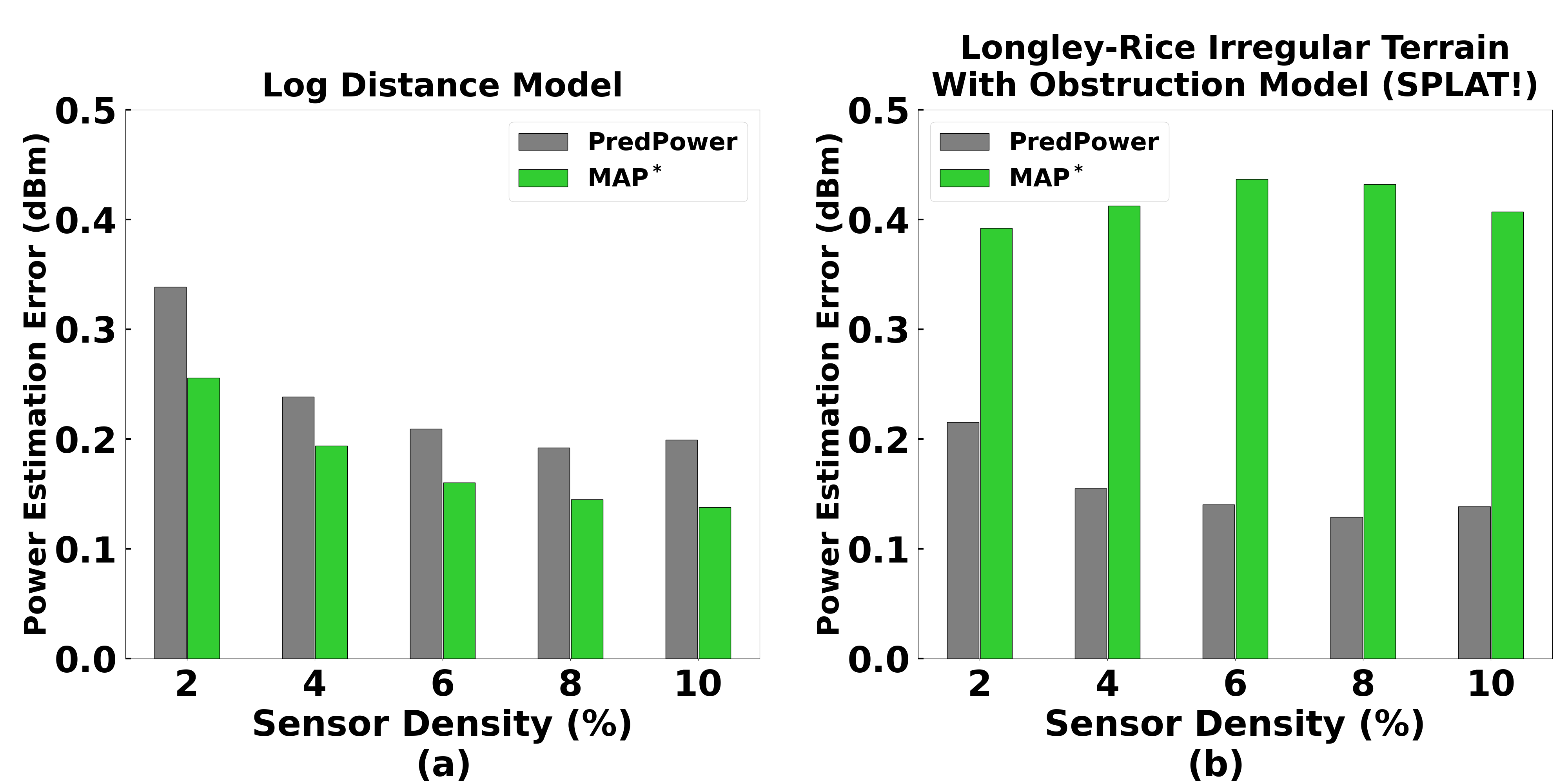}
    \caption{The single transmitter power estimation error of \power and \map in two propagation models, (a) Log-distance model and (b) Longley--Rice Irregular Terrain with Obstruction Model (SPLAT!), for varying sensor densities.}
    \label{fig:singleTXpower}
\end{figure}

Figure~\ref{fig:singleTXpower}(a) shows the performance of single transmitter power estimation in the log-distance propagation model scenario with varying sensor density.
In this case, \map has a 10 to 20 percent smaller power estimation error.
Figure~\ref{fig:singleTXpower}(b) shows the performance of single transmitter power estimation in the SPLAT! model with varying sensor density.
In this case, \power is significantly lower in power error.
So in average, \power outperforms \map in single transmitter power estimation.
We can also conclude that for \power, a higher sensor density will decrease the power estimation error. \
While a 2\% of sensor density will lead to a higher error, a sensor density of 6\% is enough to give relatively good results.

For multiple transmitter power estimation, we compare three methods in two propagation models and show that \power with error correction has the best performance among the three methods.
\power without error correction is expected to perform the worst and it suggests that the post-processing error correction stage for \power is important and works well.
\begin{figure}[t]
    \centering
    \includegraphics[width=0.75\textwidth]{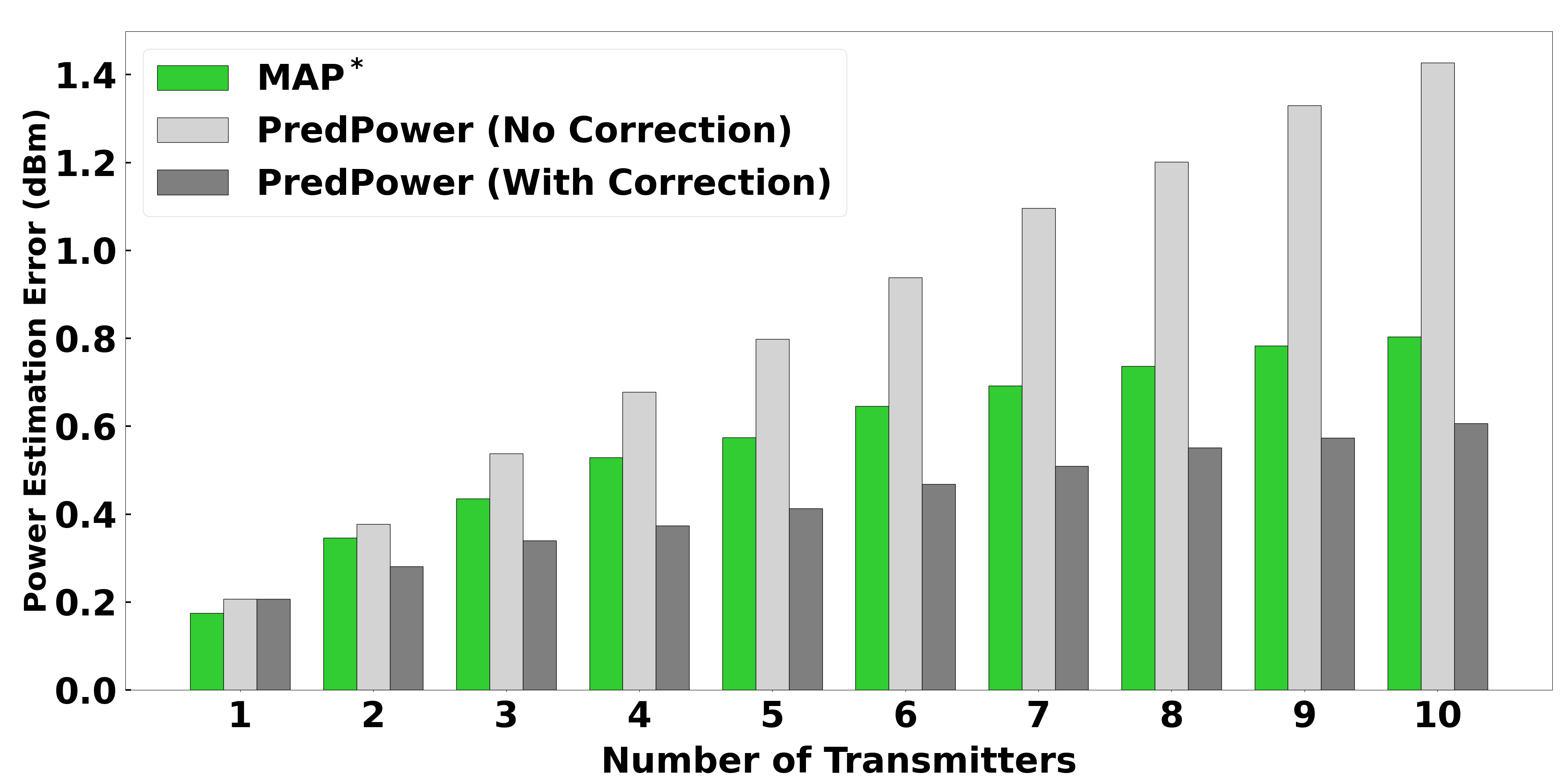}
    \caption{The transmitter power estimation error of \map, \power with and without correction in Log-distance model for varying number of intruders}
    \label{fig:logdistance-multiTXpower}
\end{figure}
\begin{figure}[t]
    \centering
    \includegraphics[width=0.75\textwidth]{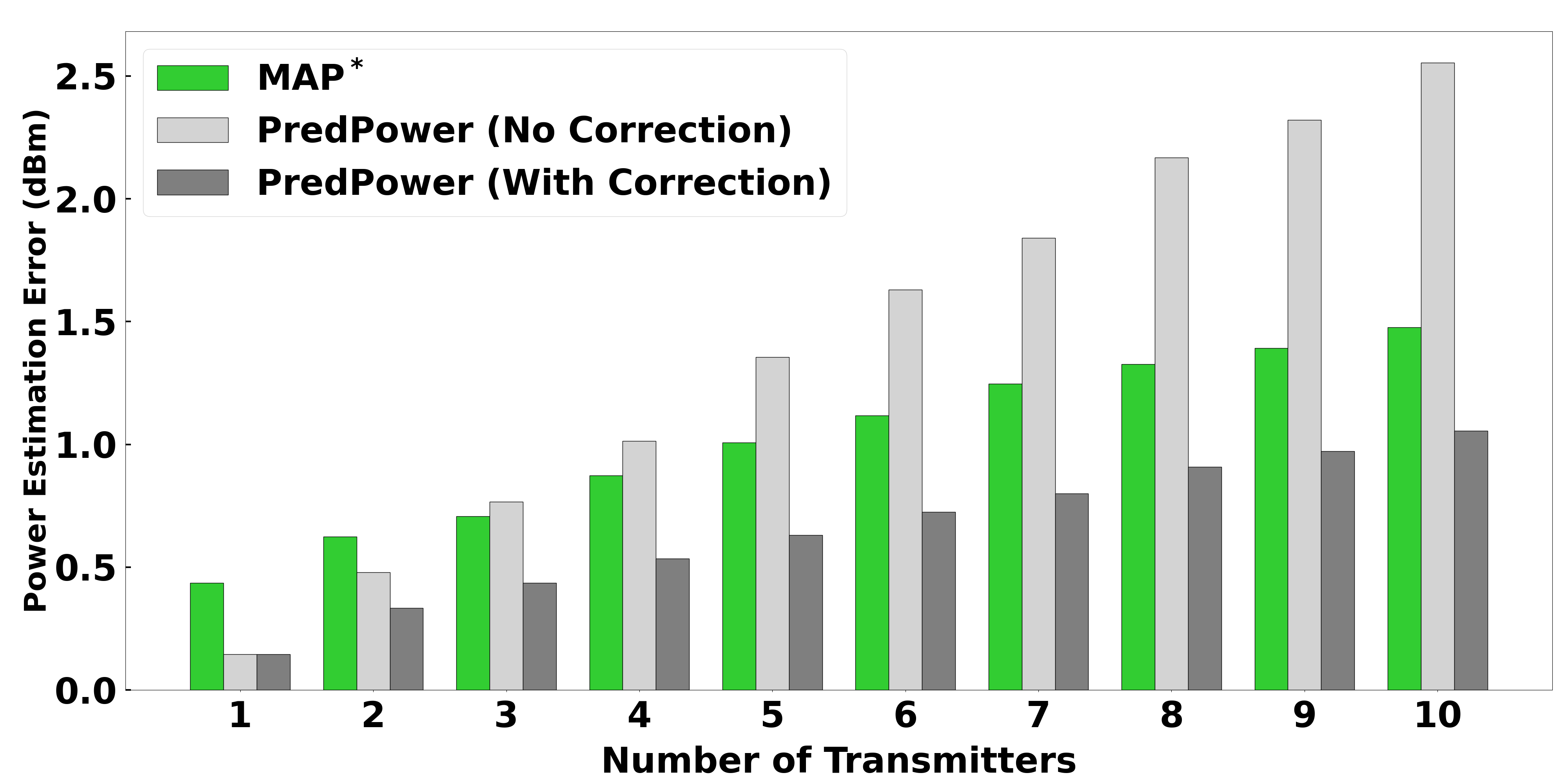}
    \vspace{-0.1in}
    \caption{The transmitter power estimation error of \map,  \power with and without correction in Longley--Rice Irregular Terrain with Obstruction Model (SPLAT!) for varying number of intruders.}
    \label{fig:splat-multiTXpower}
\end{figure}
Figure~\ref{fig:logdistance-multiTXpower} shows the power estimation error of three methods with a varying number of transmitters while the sensor density is 6\%.
In this figure, \map is the best only when the number of transmitters is one (which is consistent with Fig~\ref{fig:singleTXpower}(a)).
Also the number of transmitters is one is the only case when \power with correction and without correction has the same performance.
This is also expected because there is no need to error correction when there is only one transmitter in the area.
In all other cases, we see that \power with error correction is the best, \power without error correction  is the worst, and \map is in the middle.
In Figure~\ref{fig:splat-multiTXpower}, which shows experiment results running in the SPLAT! propagation model, we see a similar pattern compared to Figure~\ref{fig:logdistance-multiTXpower}.
The difference is that \power with error correction is always the best and the power error is larger than the log-distance model scenario.
For example in Figure~\ref{fig:logdistance-multiTXpower}, the power estimation error for \power with error correction goes up to 0.6 dB, where as in Figure~\ref{fig:splat-multiTXpower}, the error goes up to 1 dB.

\subsection{\bf Evaluation over Testbed Data}
\label{subsec:ipsn}
In this subsection, we show that our \our performs well in real-world collected data.
For this, we repurpose our testbed data from~\cite{ipsn20-mtl} as described below. We start with describing our testbed data from~\cite{ipsn20-mtl}.

\para{Testbed Data.}
In \cite{ipsn20-mtl}, we conducted a testbed in an outdoor parking area of $32m\times 32m$ large.\footnote{Dataset publicly available at: \url{https://github.com/Wings-Lab/IPSN-2020-data}}
Each grid cell has a size of $3.2m \times 3.2m$, with the grid size being $10\times 10$.
We place a total of 18 sensors on the ground.
The sensors consist of Odroid-C2 (a single-board computer) connected to an RTL-SDR dongle and the RTL-SDR connects to dipole antennas.
The transmitters are USRP or HackRF connecting to a laptop.
We collect raw Inphase-Quadrature (I/Q) samples from the RTL-SDR at the 915 MHz ISM band.
We perform FFT on the I/Q samples with a bin size of 256 samples to get the signal power values, and then utilize the mean and standard deviation of the power at frequency 915 MHz reported from each of the sensors.

\begin{figure}[t]
    \centering
    \includegraphics[width=0.98\textwidth]{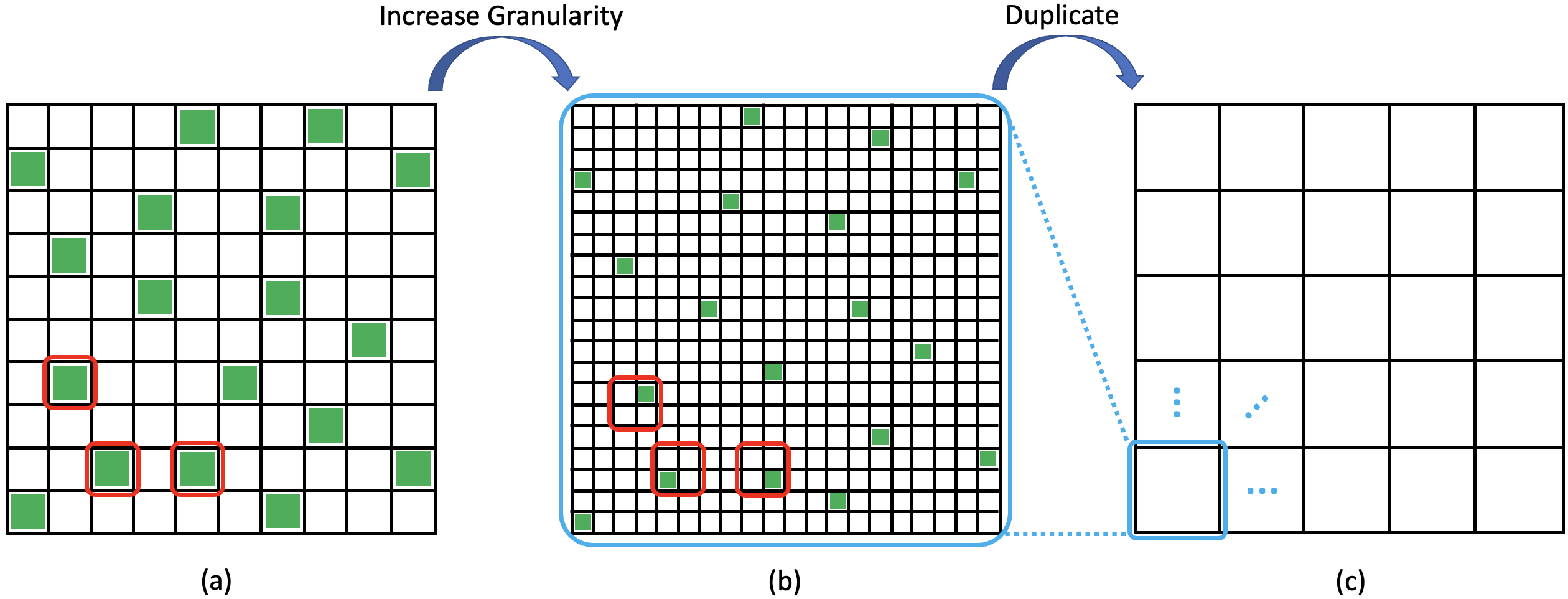}
    \caption{(a). The original $10\times10$ testbed grid with 18 sensors (green cells) representing a $32m \times 32m$ area. (b). The $20\times20$ grid (a tile) obtained by replacing each original cell by $2 \times 2$ smaller cells; a sensor, if present in the original cell, is placed in a random cell within the  $2\times2$ grid (see the green cells). (c). The final $100\times100$ grid obtained by duplicating the $20\times20$ tile 25 times using a $5\times5$ pattern. The final geographic area is $160m\times160m$.}
    \label{fig:testbed}
\end{figure}

\para{Transforming the Data from $10 \times 10$ to $100 \times 100$ Grid.}
Note that \our's input requires a $100\times 100$ input, while the above data is over a
 $10\times 10$ grid. Also, the sensor density in the above data is 18\%, while we desire a sensor
 density of around 4-6\% to have a fair comparison with our simulation based evaluations in previous subsections. To achieve these objectives, we transform the above $10 \times 10$ data to a $100 \times 100$ grid data in two steps as follows.
\begin{enumerate}
    \item Increase the data granularity from $10 \times 10$ to $20\times 20$, by dividing each cell into $2 \times 2$ cells; we randomly pick one of these four smaller cells to represent the original cell (i.e., to place the sensor if it existed in the original cell). See the red-bordered boxes in Fig.~\ref{fig:testbed}(a)-(b). We refer to the full $20 \times 20$ grid as a \emph{tile}.
    \item Now, we duplicate the $20\times20$ tile $25$ times using a $5 \times 5$ pattern to generate a  $100 \times 100$ grid. See Fig.~\ref{fig:testbed}(b)-(c).
\end{enumerate}
The above steps effectively increase the area from the original $32m\times32m$ to $160m\times160m$. 
Note that the first step above only splits each original cell into four smaller cells without increasing the whole area size.
The $100\times100$ grid will have a sensor density of 4.5\% and each grid cell represents an area of $1.6m\times1.6m$.


We note that the second duplication step can introduce inaccurate sensor readings at the tile's ``edges", due to 
any transmitters from adjoining tiles. To circumvent this issue, we place {\em transmitters} only within the internal
$10 \times 10$ cells of each $20 \times 20$ tile (i.e., avoid placing a transmitter on the five-cell edge of each tile). This
yields a total of 2500 potential positions to place a transmitter in the final $100 \times 100$ grid. 
With the above setting, we generate training and testing datasets consisting of 25,000 and 12,500 samples respectively.

\begin{figure}
    \centering
    \includegraphics[width=0.95\textwidth]{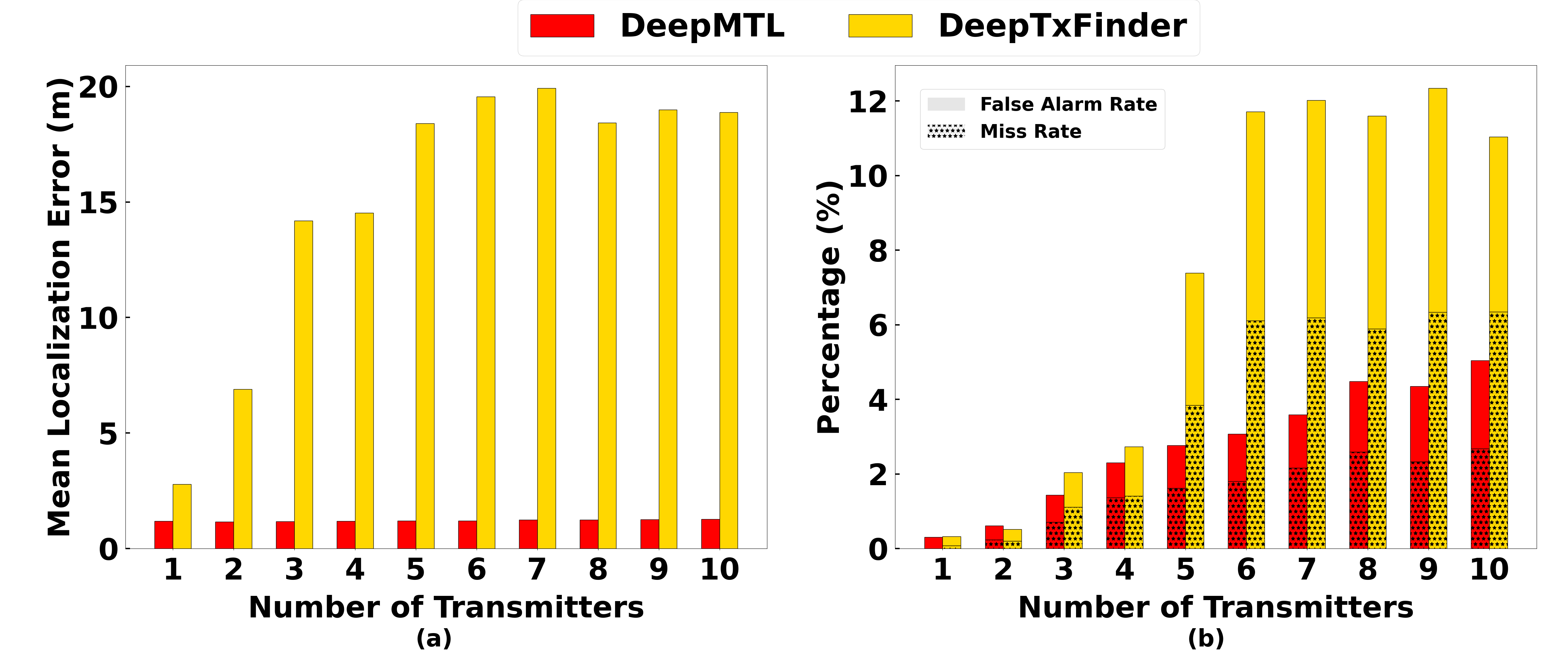}
    \vspace{-0.1in}
    \caption{The localization error (a), false alarm rate and miss rate (b) of \our and \deeptx in a real world collected data for varying number of intruders.}
    \label{fig:ipsn}
\end{figure}

\para{Testbed Results.}
The performance of \our on this real world based data is shown in Fig.~\ref{fig:ipsn}.
Compared to \deeptx, \our is significantly better in localization error and false alarm rate and miss rate in almost all cases, which aligns to the results in the previous subsections based on data generated from either log-distance model or SPLAT!.
The localization error of \our in Fig.~\ref{fig:ipsn}(a) is around 1.3 meters.
The error increases mildly with the increase in the number of transmitters.
The localization error in the testbed data is smaller compared to both log-distance data results (Fig.~\ref{fig:logdist-error-vary_numintru}) and SPLAT! data results (Fig.~\ref{fig:splat-error-vary_numintru}).
This is because a grid cell here is representing a smaller area.
In the log-distance data, the localization error is roughly one-fourth the side length of the grid cell. 
In the SPLAT! data result, the localization error is roughly half the side length of its grid length.
In the testbed data, the localization is roughly eighty percent the side length of a grid cell.
So the localization error in the testbed data is the highest relative to the length of a grid cell it represents.
The sum of false alarm rate and miss rate is 3\% when the number of transmitters is five and is 5\% when the number of transmitters is ten.
The results are a little bit worse than the results in the SPLAT! data (Fig.~\ref{fig:splat-missfalse-vary-numintru}), where the sum is 2\% for five transmitters and 4\% for ten transmitters.

\section{\bf Related Work}
\label{sec:related}
\para{Spectrum sensing} is usually being realized by some distributed crowdsourced low-cost sensors. 
Electrosense~\cite{electrosense} and its follow up work Skysense \cite{mobisys20-skysense} are typical work of spectrum sensing.
In this crowdsourced sensing paradigm~\cite{chakraborty2017specsense}, sensors collect I/Q samples (in-phase and quadrature components of raw signals) and compute PSD (power spectral density), which is RSS.
Crowdsourced low-cost sensors do not have the capability to collect AoA (angle of arrival) data because it requires more expensive antenna arrays.
They also do not have the capability to collect ToA (time of arrival) data because it requires the transmission of a wide-band known sequence~\cite{pimrc2021-localize}, which is impossible in the case of localizing (blind) intruders.
Spectrum sensing platforms serve as the foundation of the spectrum applications, and transmitter localization is one of the main applications.
Other applications include signal classification~\cite{toccn18-sigclassify}, spectrum anomaly detection~\cite{ben-zhao}, sensor selection~\cite{ton-sensorselect,bhattacharya2022fast}, spectral occupancy estimation~\cite{mobicom21-deepradar}, etc.

\para{Transmitter localization.} Localization of an intruder in a field using sensor observations has been widely studied, but most of the works have focused on
localization of a single intruder~\cite{our-infocom,dutta2016see}.
In general, to localize multiple intruders, the main challenge comes
from the need to ``separate'' powers at the sensors~\cite{mobicom-30},
i.e., to divide the total received power into power received from
individual intruders. Blind source separation is a very challenging
problem; only very limited settings allow for known
techniques using sophisticated receivers~\cite{freq-sig,ben-zhao}.
We note that (indoor) localization of a
  device~\cite{infocom00-radar} based on signals received from multiple reference points (e.g, WiFi access
  points) is a quite different problem
  (see~\cite{zafari-19} for a recent survey), as the signals from
  reference points remain separate, and localization or tracking of multiple
  devices can be done independently.
  Recent works on multi-target localization/tracking such as~\cite{ipsn19-multipassive} are different in the way that targets are passive, instead of active transmitters in the \mtl problem.
Among other related works,~\cite{multi-tx-dyspan-19} addresses the challenge of handling time-skewed sensors observations in the MTL problem.
\eat{In absence of blind separation methods, to the best of our knowledge,
only a few works have addressed multiple intruder(s) localization. 
In particular,
(i)\splot~\cite{mobicom17-splot} decomposes the multi-transmitter
localization problem to multiple single-transmitter localization
problems based on the sensors with highest of readings in a
neighborhood, (ii)~\cite{clustering} uses a clustering-based approach, (iii)~\cite{Quasi-EM} uses an
EM-based approach, (iv) \map~\cite{ipsn20-mtl}, uses 
a hypotheses-based Bayesian
approach in conjunction with a divide-and-conquer strategy to first localize 
``isolated"  intruders and then localize the remaining intruders,
(v) 
We note that the techniques of~\cite{mobicom17-splot,Quasi-EM} assume a propagation
model, while that of~\cite{clustering,Quasi-EM} require a priori
knowledge of the number of intruders present. 
Schemes in~\cite{Quasi-EM} and~\cite{clustering} have been shown inferior in performance in~\cite{mobicom17-splot,ipsn20-mtl}.
}

\para{Wireless localization} techniques mainly fall into two categories: geometry mapping and fingerprinting-based.
Geometry mapping mainly has two subcategories: ranging-based such as trilateration (via RSS/RSSI, ToA, TDoA) and direction-based such as triangulation (via AoA).
Fingerprinting-based methods can use all signal physical measurements including but not limited to amplitude, RSS/RSSI, ToA, TDoA, and AoA.
Whenever deep learning is used for localization, it can be considered as a fingerprinting-based method, since it requires a training phase to survey the area of interest and a testing phase to search for (predict) the most likely location.

\para{Deep learning for wireless localization}. 
Quite a few recent works have harnessed the power of deep learning in the general topic of localization.
E.g., DeepFi in~\cite{DeepFi2016} designs a restricted Bolzmann machine that localizes a single target using WiFi CSI amplitude data. 
DLoc in~\cite{mobicom20-deeploc} uses WiFi CSI data as well. 
Its novelty is to transform CSI data into an image and then uses an image-to-image translation method to localize a single target.~MonoDCell in~\cite{sigspatial19-monodcell} designs an LSTM that localizes a single target in indoor environment using cellular RSS data.~\cite{pimrc2021-localize} designs a three-layer neural network that locations a single transmitter.~\deeptx in~\cite{icccn20-deeptxfinder} uses CNN to address the same \mtl problem using RSS data in this paper.

\para{Transmitter Power Estimation.} There are several works that estimate the transmission power of a single transmitter.~\cite{PowerEstimate2010Zafer} studies the ``blind" estimation of transmission power based on received-power measurements at multiple cooperative sensor nodes using maximum likelihood estimation. Blind means there is no prior knowledge of the location of the transmitter or transmit power.~\cite{Ureten2011powerlocation} propose an iterative technique that jointly estimate the location and power of a single primary transmitter.
In~\cite{icoin2007-powerposition}, the primary transmitter location and power is jointly estimated by a constrained optimization method.~\cite{ipsn20-mtl} uses the maximum likelihood estimation method to estimate the power of an isolated single transmitter and adopts an online learning method to estimate the power of multiple close by transmitters simultaneously.

\section{Conclusion}

In this paper, we have designed and developed some novel deep-learning based scheme (\our) for
the multiple transmitter localization (\mtl) problem.
We extended this problem to localizing the intruders in the presence of authorized users and developed a novel technique to solve it. 
We also developed a novel technique that can solve the multiple transmitter power estimation (\mtpe) problem.
Solving the general \mtl and \mtpe are both achieved by utilizing our robust \our as a building block. 
We evaluated all our methods extensively through data simulated from two propagation models as well as a small-scale data collected from a real world testbed. 
Our developed technique outperforms prior approaches by a significant margin in all performance metrics.


\section{Acknowledgements}
This work is supported by National Science Foundation grants CNS-1642965, CNS-1815306, and CNS-2128187. The authors would like to thank the reviewers for the valuable feedback and advice.

\bibliographystyle{elsarticle-num}
\bibliography{bib}

\end{document}